\documentclass[epj,nopacs]{svjourx}
\usepackage{graphicx}
\def\lsim{\raise0.3ex\hbox{$<$\kern-0.75em\raise-1.1ex\hbox{$\sim$}}}
\def\gsim{\raise0.3ex\hbox{$>$\kern-0.75em\raise-1.1ex\hbox{$\sim$}}}

\begin{document}

\title{Heavy-flavour spectra in high energy nucleus-nucleus collisions} 

\author{W.M. Alberico\inst{1,2} \and A. Beraudo\inst{1,2,3,4} \and 
A. De Pace\inst{2} \and A. Molinari\inst{1,2} \and M. Monteno\inst{2} \and 
M. Nardi\inst{2} \and F. Prino\inst{2}}

\institute{Dipartimento di Fisica Teorica dell'Universit\`a di Torino,\\
via P.Giuria 1, I-10125 Torino, Italy 
\and
Istituto Nazionale di Fisica Nucleare, Sezione di Torino, \\ 
  via P.Giuria 1, I-10125 Torino, Italy 
\and
Centro Studi e Ricerche \emph{Enrico Fermi}, Piazza del Viminale 1, Roma, Italy
\and
Physics Department, Theory Unit, CERN, CH-1211 Gen\`eve 23, Switzerland
}

\date{}

\abstract{ 
The propagation of the heavy quarks produced in relativistic nucleus-nucleus
collisions at RHIC and LHC is studied within the framework of Langevin
dynamics in the background of an expanding deconfined medium described by
ideal and viscous hydrodynamics. The transport coefficients entering into the
relativistic Langevin equation are evaluated by matching the hard-thermal-loop
result for soft collisions with a perturbative QCD calculation for hard
scatterings. The heavy-quark spectra thus obtained are employed to compute the
differential cross sections, the nuclear modification factors $R_{AA}$ and the 
elliptic flow coefficients $v_2$ of electrons from heavy-flavour decay.}

\maketitle

\section{Introduction}

The aim of the ongoing heavy ion programme at RHIC (with Au-Au collisions at
$\sqrt{s}_{\rm NN}$ up to 200 GeV) and of the recently started programme at LHC 
(with the first Pb-Pb collisions at $\sqrt{s}_{\rm NN}=2.76$~TeV) is the 
creation of a deconfined system of quarks and gluons, with a lifetime 
sufficiently long (a few fm) to allow measurable signals to show up.
The challenge is not only to establish the formation of the deconfined phase, 
but also to study the properties of the produced medium. 
Among the most interesting observables investigated at RHIC, perhaps, there are 
the quenching of high-$p_T$ hadron spectra, expressed through the 
\emph{nuclear modification factor} $R_{\rm AA}(p_T)$~\cite{RAAphe,RAAsta}, and 
the azimuthal anisotropy of particle emission in the transverse plane,
quantified by the \emph{elliptic-flow coefficient} $v_2$~\cite{v2phe,v2sta}. 
The asymptotic value $R_{\rm AA}(p_T)\approx 0.2$ found at high $p_T$ indicates 
that the matter created in the heavy ion collisions is very opaque, leading to
a sizable energy loss of the few high-momentum partons produced in the initial
instants of the collisions by hard processes. The latter are describable in the
framework of perturbative Quantum Chromodynamics (pQCD).
Furthermore, the elliptic flow observed in semi-central collisions seems to find
a natural explanation within a hydrodynamical description of the medium,
displaying an expansion driven by the pressure gradient: the latter, resulting
larger along the reaction plane, leads to the azimuthal anisotropy of the
transverse-momentum spectra. 
Notably, the large value of $v_2$ found in the data, besides supporting the
hydrodynamical behavior of the created matter (entailing a mean free path
$\lambda_{\rm mfp}$ much smaller then the system size $L$), requires also a
quite short thermalization time ($\tau_0\lsim1$ fm). 
$R_{AA}$ and $v_2$ have recently been measured also by the ALICE detector at
LHC for unidentified particles~\cite{v2ali,RAAali}.

These two observables ($R_{\rm AA}$ an $v_2$) have been initially studied 
for the case of light-flavour hadrons (pions, kaons, protons, ...), arising from 
the hadronization of gluons and light quarks. 
In this context, the quenching of hadron spectra has been attributed to the
energy-loss of hard partons due to the radiation of soft gluons ($\omega\ll
E_{\rm hard}$) occurring in the presence of scattering centers (the plasma
particles). Different theoretical approaches have been 
proposed~\cite{glv,bdmps,asw,amy}, all stressing the importance of coherence
(Landau-Pomeranchuk \cite{lpm1} and Migdal \cite{lpm2}) effects for large
enough formation times of the emitted gluon, so that 
$\tau_{\rm form}\gsim\lambda_{\rm mfp}$. 

If the soft-gluon radiation were the only mechanism responsible for energy loss,
one would expect heavy quarks ($c$ and $b$) to be much less quenched.
Characteristic effects occurring for massive particles are 
the suppression of collinear gluon emission for angles $\theta\lsim M/E$
(\emph{dead cone effect}~\cite{dead1,dead2}), the shortening of the gluon
formation time and the suppression of the radiated gluon spectra at large
energy \cite{Arm}.
Furthermore, while light-hadron spectra get contribution from the fragmentation
of partons belonging both to the fundamental (quarks) and the adjoint (gluons)
representations, heavy-flavour electrons arise only from the decays of
$c$ and $b$ quarks. Hence, the associated Casimir factor $C_F$ (smaller then 
$C_A$ of the gluons) should entail a larger mean-free-path, resulting in a
lower rate of gluon radiation. 
However, single-electron spectra (resulting from the semi-leptonic decays of $D$
and $B$ mesons) measured by the PHENIX experiment at RHIC show a similar level of
suppression as the one found for light hadrons~\cite{phenix,phenix2} (a similar 
analysis is being performed by the STAR collaboration); moreover, 
they also display a sizable elliptic flow. 
These findings hint at reconsidering the importance of collisional energy
loss. Calculations of the stopping power of the quark-gluon plasma (QGP) for
heavy quarks due to elastic collisions can be found for instance in
Refs.~\cite{tho12,pei1,pei2}. 

An appropriate tool to study the final spectra of heavy quarks, after their
evolution in the fireball created in the heavy ion collisions, is given by the
(relativistic) Langevin equation, which relies on the picture of many 
uncorrelated random collisions~\cite{svet,ber}. 
This approach --- pursued by solving either the Langevin equation or the
corresponding Fokker-Planck equation --- has already been explored in the
literature~\cite{tea,rapp,aic,hira,das,aic2} and will be the one addressed in
this paper. 
Solving the Langevin equation requires as an input the knowledge of a few
transport coefficients, accounting for the interaction of the heavy quarks with
the medium: indeed, in this framework, the latter gives rise to a friction
force and a momentum broadening. The various calculations available in the
literature \cite{tea,rapp,aic,hira,das,aic2} differ mainly in the way they
estimate the above coefficients, either inspired by pQCD or within less
conventional scenarios (assuming for instance the existence of resonant
states in the plasma~\cite{rapp} or the AdS/CFT correspondence~\cite{hira}). In
all the cases the freedom of tuning some parameter allows one to explore
more/less strongly coupled regimes. 

However, before looking for alternative explanations, it appears important to
have at hand a fully consistent weakly coupled calculation, to check how closely
the experimental data can be reproduced within a perturbative picture and
how much room is left for more ``exotic'' scenarios. 
This is the major goal of our paper, in which we provide a microscopic
evaluation of the heavy-quark transport coefficients and we use them to solve
the Langevin equation for $c$ and $b$ quarks that are produced in nucleus-nucleus
collisions and are let propagate randomly in the resulting deconfined medium. 
Our calculation relies on the separate treatment of soft and hard
collisions. The first ones, mediated by the exchange of long wavelength gluons,
will be described within the Hard Thermal Loop (HTL) approximation, which
resums medium effects. Hard collisions, involving a high-momentum
exchange, will be calculated in kinetic theory, employing leading order pQCD
matrix elements~\cite{com}. The two contributions will then be summed up. 
Once the transport coefficients have been calculated, we can address the
numerical solution of the Langevin equation, given suitable initial conditions
and a background medium.

The initial heavy-quark pairs are produced using the POWHEG package
\cite{nason1,nason2}, 
a hard event generator for heavy-quark production in hadronic collisions, which
implements pQCD at next-to-leading order (NLO). Since it generates events, this
code is particularly suitable for a Langevin simulation.
Care is taken to allow also for shadowing and transverse momentum broadening
effects. 
Once created, the heavy quark is allowed to propagate in the hot medium
according to the Langevin dynamics, with transport coefficients calculated
as mentioned above. 
The evolution of the background medium is described by two different scenarios,
namely ideal \cite{kolb1,kolb2,kolb3} and viscous \cite{rom1,rom2,rom3}
hydrodynamics, in order to estimate the dependence of our results on the
hydrodynamical scheme.
Around the phase transition energy density the $c$ and $b$ quarks are made 
hadronize (at the moment, for the sake of simplicity, only through
fragmentation) and then decay to electrons, to allow for a comparison with the
existing experimental data from RHIC for open heavy-flavour electrons. 
Results will 
be presented for the invariant yields, the nuclear modification factors and the 
elliptic flow coefficients at the kinematics of both RHIC and LHC. 
For the latter we have chosen the highest center-of-mass energy that the heavy
ion experiments should be able to reach, i.~e. $\sqrt{s}=5.5$~TeV. Preliminary
results have already been presented in Refs.~\cite{bers}. 

Our paper is organized as follows: in Sect.~\ref{sec:Theo} we introduce the
theoretical tools employed in generating the heavy-quark spectra after a
high-energy heavy ion collision, examining all the steps of the process, from
the production stage, through the hydrodynamic evolution, the Langevin
propagation, the calculation of the transport coefficients, up to hadronization
and decay; in Sect.~\ref{sec:res} we present and discuss our findings for the
differential spectra, the nuclear modification factor and the elliptic-flow
coefficient; finally, in Sect.~\ref{sec:concl} we sum up our conclusions; some
more technical material is deferred to the Appendices.

\section{Theoretical framework}
\label{sec:Theo}

\subsection{Heavy-quark production in $pp$ and $AA$ collisions}
\label{subsec:ini}

For every experimental setup (either $pp$ or $AA$ at $\sqrt{s}=200$~GeV (RHIC)
and $\sqrt{s}=5.5$~TeV (LHC)) we have generated $4.5\times10^7$ heavy-quark
pairs (either $c\bar{c}$ or $b\bar{b}$) using the POWHEG event generator
\cite{nason1,nason2}, which implements pQCD at NLO. 

\begin{table}
\caption{\label{tab:qqbarcs} Total production cross sections for heavy-quark
  pairs calculated with POWHEG for $pp$ collisions at two different 
  center-of-mass energies and for two choices of PDF's: CTEQ6M alone (suitable
  for bare $pp$ collisions) and supplemented by the EPS09 nuclear
  modifications \cite{eps} (for NN events embedded in a nucleus-nucleus
  collision).
}
\begin{center}
\setlength{\tabcolsep}{4pt}
\begin{tabular}{|c|cc|cc|}
\hline
  & \multicolumn{2}{c|}{$\sigma_{c\bar{c}}$ ($mb$) per NN event} & 
    \multicolumn{2}{c|}{$\sigma_{b\bar{b}}$ ($mb$) per NN event} \\
\hline
  $\sqrt{s}$ & CTEQ6M & (+EPS09) & CTEQ6M & (+EPS09) \\
\hline
  $200$ GeV  & 0.254 & 0.236 & 1.77$\times 10^{-3}$ & 
    2.03$\times 10^{-3}$ \\
  $5.5$ TeV  & 3.015 & 2.288 & 0.187 & 0.169 \\
\hline
\end{tabular}
\end{center}
\end{table}
There is considerable uncertainty in the $Q\bar{Q}$ production cross section,
due to uncertainties in the quark masses and in the renormalization and
factorization scales. Here we have chosen $m_c=1.5$~GeV, $m_b=4.8$~GeV and for
the values of the renormalization and factorization scales we have kept the
default choice in POWHEG, that is the heavy-quark transverse mass in the
$Q\bar{Q}$ reference frame \cite{nason1}.
For $pp$ events we have employed the CTEQ6M parton distribution function
(PDF). The total $c\bar{c}$ and $b\bar{b}$ production cross sections one gets
at RHIC energies (see Table~\ref{tab:qqbarcs}) are very close to the central
values predicted by FONLL \cite{cacc}.
Another important effect to be accounted for in the production cross sections is
related to the intrinsic transverse momentum of the initial partons. We have
adopted the procedure described in Ref.~\cite{vogt}, which amounts to add to
the out-coming heavy quarks a transverse momentum contribution randomly
generated from a Gaussian distribution with given variance $\langle
k_T^2\rangle_{\rm NN}$. For $pp$ collisions we have adopted the standard value
$\langle k_T^2\rangle_{\rm NN}=1 {\rm ~GeV}^2/{\rm c}^2$.

\begin{figure}
\includegraphics[clip,width=0.48\textwidth]{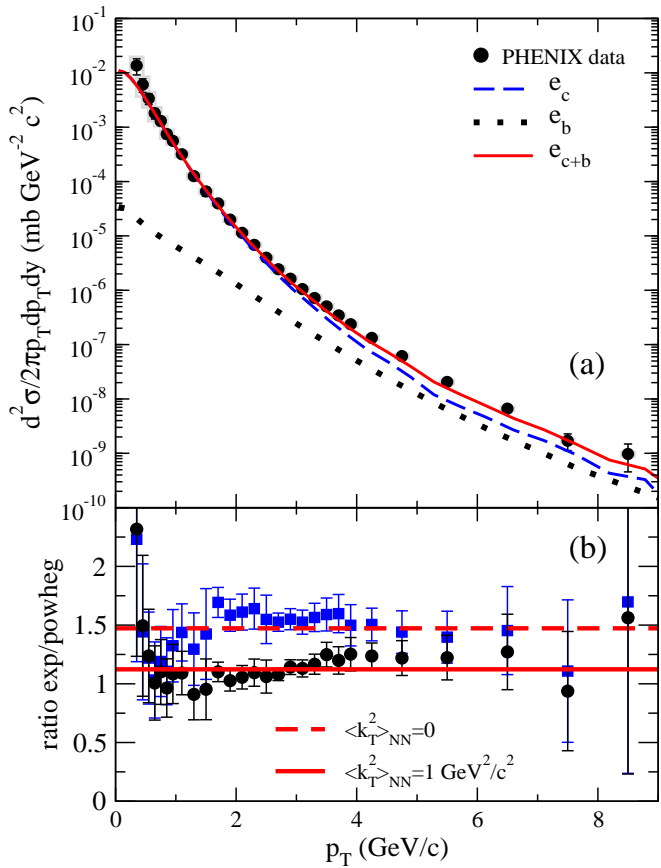}
\caption{(a) Invariant differential cross section of electrons [$(e^++e^-)/2$],
  from heavy-flavour decay in $pp$ collisions at $\sqrt{s}=200$~GeV, at
  mid-rapidity as a function of transverse momentum. The experimental points
  represent PHENIX data~\protect\cite{phenixpp,phenix2}; both statistical 
  (error bars) and systematic (gray boxes) errors are displayed. 
  The curves represent the contributions coming from electrons
  originating from a charm or bottom quark generated by POWHEG with inclusion of
  transverse broadening. 
  (b) ratio of data to the POWHEG prediction as a function of $p_T$, with
  (circle) and without (square) inclusion of transverse broadening; the lines
  represent the best fit to a constant.}
\label{fig:idcs_pp_RHIC}
\end{figure}
\begin{figure}
\includegraphics[clip,width=0.48\textwidth]{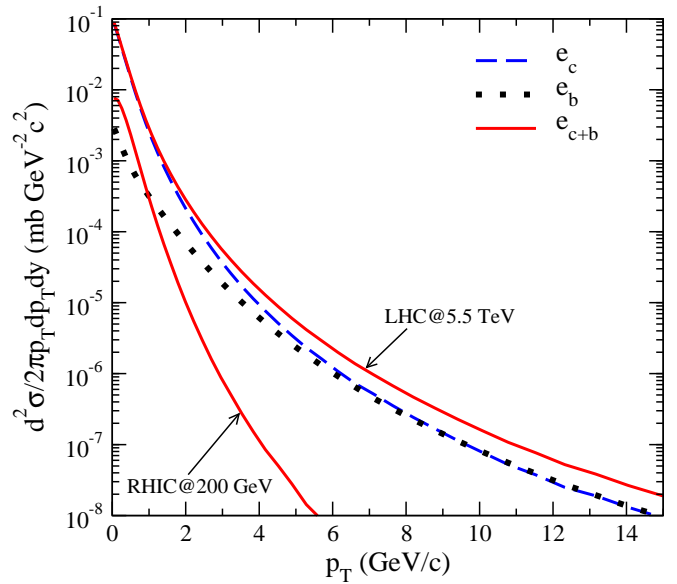}
\caption{Invariant differential cross section of electrons [$(e^++e^-)/2$],
  from heavy-flavour decay in $pp$ collisions at $\sqrt{s}=5.5$~TeV, at
  mid-rapidity as a function of transverse momentum. 
  The curves represent the contributions coming from electrons
  originating from a charm or bottom quark generated by POWHEG with inclusion of
  transverse broadening. For comparison we have also included the total yield of
  electrons at RHIC from Fig.~\protect\ref{fig:idcs_pp_RHIC}.}
\label{fig:idcs_pp_LHC5}
\end{figure}
In Fig.~\ref{fig:idcs_pp_RHIC} one can see the invariant differential cross
sections of electrons from heavy-flavour decay (see
Sect.~\ref{subsec:hadronize} for details about hadronization and decay) in $pp$
collisions at $\sqrt{s}=200$~GeV, compared to the mid-rapidity data from PHENIX
\cite{phenixpp,phenix2}\footnote{
  For the purpose of comparing to the PHENIX data, the doubly differential cross
  sections have been calculated by integrating over the experimental acceptance
  in pseudo-rapidity ($|\eta|<0.35$) and dividing by the rapidity range ($\Delta
  y\cong0.7$). For consistency, also the cross sections at the energy of LHC
  have been calculated in the same way, using the constraints of the ALICE
  experiment ($|\eta|<0.9$ and $\Delta y\cong1.8$).}. 
One observes that the pQCD outcome provides a fairly good description of both the
shape and the absolute magnitude of the data. In panel (b) of the figure the
ratio between experiment and theory is also displayed: this ratio is nearly
$p_T$-independent and a best fit with a constant provides for the latter the
value 1.12. 
Given a 10\% normalization uncertainty in the data (not shown in the figure) and
the previously mentioned uncertainties on the theoretical parameters, a 12\%
difference between theory and data can be easily accommodated. Accordingly, 
for sake of simplicity, in the following we have chosen to multiply by this
factor all the calculated cross sections at RHIC energies (of course, $R_{AA}$
and $v_2$ are not affected by cross section normalization). 
In Fig.~\ref{fig:idcs_pp_RHIC}b we also display the ratio of data to the cross
section generated without any transverse momentum broadening: in this case the
discrepancy is fairly larger (although still within the experimental and
theoretical uncertainties). 

In Fig.~\ref{fig:idcs_pp_LHC5} the invariant differential cross sections of
electrons from heavy-flavour decay in $pp$ collisions at $\sqrt{s}=5.5$~TeV are
reported: apart from the much larger strength and the less steep slope of the
cross sections, one observes that the bottom relative contribution is much more
important than at the RHIC energy.

In the case of nucleus-nucleus collisions, there are two important differences
one has to consider in connection to the initial heavy-quark distributions.
First of all, the nuclear PDF's should be different from the ones employed in
$pp$ collisions and, to account for this occurrence, we have adopted here the
EPS09 scheme~\cite{eps}. 
In principle, the density probed by the colliding partons should depend on
the impact parameter $b$: in describing nucleus-nucleus collisions we have made
the simple choice of employing the EPS09 scheme for impact parameters $b<2R$ and
of neglecting nuclear corrections for $b>2R$ ($R$ being the radius of the
nuclear density distribution).
The main consequence of using a different PDF in $AA$ and $pp$ collisions
relates to the different total (per binary collision) $c\bar{c}$ and $b\bar{b}$
production cross sections one gets, as one can see in Table~\ref{tab:qqbarcs}.
Indeed, since in a binary process the longitudinal momentum fraction carried by
the two initial partons (most of the times gluons) is given by \cite{PPR}
\begin{equation}
x_{1/2}=M_{Q\bar{Q}}/\sqrt{s}_{\rm NN}e^{\pm y_{Q\bar{Q}}},
\end{equation}
$M_{Q\bar{Q}}$ and $y_{Q\bar{Q}}$ being the invariant mass and rapidity of the
$Q\bar{Q}$ pair, one can get larger or smaller cross sections in $AA$
collisions, at different energies and quark masses, depending upon whether one is
probing the anti-shadowing or shadowing regions.
For instance, at RHIC energy, because of the large bottom mass, the main
contributions come from relatively large values of $x$, where anti-shadowing
dominates --- hence a larger $b\bar{b}$ cross section per binary collision
results --- whereas at LHC energies one is mainly probing the low $x$ shadowing
region, yielding smaller cross sections. This fact has sizable consequences on
$R_{AA}$, as we shall see later. 

A second difference one has to cope with in $AA$ collisions concerns the larger
transverse momentum acquired on average by the colliding partons, because of
the large size of the traversed medium. To get a realistic estimate for 
$\langle k_T^2\rangle_{AA}$ in nucleus-nucleus collisions we have adopted a
Glauber approach --- developed in Refs.~\cite{huf1,huf2} in order to
study the $p_T$ distribution of charmonia produced in proton-nucleus and
nucleus-nucleus collisions \cite{abreu} --- and we have generalized it to study
the inclusive single-quark spectra. Details of the procedure can be found in
Appendix~\ref{app:kTAA}. One gets an average squared transverse momentum that
depends not only on the impact parameter of the collision and on the nuclei 
involved, but also on the transverse position of the $Q\bar{Q}$ pair. 
To get an orientation on the magnitude of the quantities involved, we mention
that at RHIC energy one gets, depending upon the impact parameter, average
values of $\langle k_T^2\rangle_{AA}$ around  
$1.3\div1.5 {\rm ~GeV}^2/{\rm c}^2$ for charm and around 
$1.8\div2.3 {\rm ~GeV}^2/{\rm c}^2$ for bottom in Au-Au collisions; at the LHC
($\sqrt{s}= 5.5$~TeV) around $1.5\div1.8 {\rm ~GeV}^2/{\rm c}^2$ for charm and
around  $2.5\div3.3 {\rm ~GeV}^2/{\rm c}^2$ for bottom in Pb-Pb collisions.

Finally, for $AA$ collisions, since we are following in the Langevin framework
the space-time propagation of an ensemble of heavy quarks, we have also to
specify their space-time distribution at the onset of the hydrodynamical
evolution of the background medium.
This we do consistently with the choice discussed below for the initial
conditions of the hydrodynamic equations. Namely, at the initial proper time
$\tau_0$ the position of the heavy quarks in the transverse plane is calculated
in a Glauber framework according to a distribution generated by the nuclear
overlap function $T(x+b/2,y)T(x-b/2,y)$, where
\begin{equation}
  T(x,y)=\int dz\,\rho(x,y,z),
\label{eq:TA}
\end{equation}
$\rho$ being a Fermi parameterization of the nuclear density \cite{DeV87}
and $b$ the impact parameter. In the longitudinal direction we set $z=\tau_0
\sinh\eta_s$, with
\begin{equation}
  \eta_s\equiv\frac{1}{2}\ln\frac{t+z}{t-z}=
    \frac{1}{2}\ln\frac{E+p_z}{E-p_z}.
\end{equation}

\subsection{Hydrodynamic evolution at RHIC and LHC}
\label{subsec:hydro}

\begin{table}[b]
\caption{\label{tab:hydro} Values of the initial proper time $\tau_0$, central
  entropy density $s_0$ and central temperature $T_0$ employed as initial
  conditions for the hydrodynamical evolution.}
\setlength{\tabcolsep}{1pt}
\begin{center}
\begin{tabular}{|c|ccc|ccc|}
\hline
  & \multicolumn{3}{c|}{$\eta/s=0$} & \multicolumn{3}{c|}{$\eta/s=0.08$} \\
\hline
  & $\tau_0$ (fm) & $s_0$ (fm$^{-3}$) & $T_0$ (MeV) 
  & $\tau_0$ (fm) & $s_0$ (fm$^{-3}$) & $T_0$ (MeV) \\
\hline
       &      &     &     & 0.1 & 840 & 666 \\
  RHIC & 0.6  & 110 & 357 & 0.6 & 140 & 387 \\
       &      &     &     & 1   & 84  & 333 \\
\hline
\hline
        & 0.1  & 2438 & 1000 & 0.1  & 1840 & 854 \\
  LHC   & 0.45 & 271  & 482  & 1    & 184  & 420 \\
\hline
\end{tabular}
\end{center}
\end{table}

Hydrodynamics has been successfully applied to the description of collective
phenomena in heavy ion collisions at RHIC, yielding a sensible description for
a number of experimental observables. For our purposes, hydrodynamics provides
the full space-time evolution of the properties of the expanding medium ---
such as  temperature, flow velocity and energy density --- that are needed to
follow the propagation of the heavy quarks. We have chosen two different
implementations of the relativistic hydrodynamic equations, whose codes are
publicly available, namely ideal \cite{kolb1,kolb2,kolb3} and viscous
\cite{rom1,rom2,rom3} hydrodynamics, both assuming exact longitudinal boost
invariance. The two models differ not only in the ideal/viscous implementation
(with the ratio of shear viscosity to entropy density taken to be
$\eta/s=0.08$ in the viscous case), but also in the choice of the equation of
state (EOS) and of the initial conditions. By comparing the results obtained in
the two scenarios one can get an estimate of the amount of uncertainty due to
the treatment of the background medium. 

The initial energy density distribution is computed in both cases according to
the Glauber model, using either the number of
participating nucleons ($N_{\rm part}$) or the number of binary collisions
($N_{\rm coll}$). In the ideal model \cite{kolb1} the distribution has been
ascribed for the 75\% to ``soft'' processes scaling as $N_{\rm part}$ and for
the 25\% to ``hard'' processes scaling as $N_{\rm coll}$; in the viscous model
the initial energy density is proportional to $N_{\rm coll}$.
Concerning the EOS, in the ideal case it is obtained by matching, through a
Maxwell construction, a gas of hadron resonances with an ideal gas of massless
quarks and gluons; this model yields a first order transition at 
$T_{\rm crit}=164$~MeV. In the viscous case, one employs the
more realistic EOS of Ref.~\cite{Lai06}, which implements a crossover between
a low temperature hadron-resonance gas and a high temperature pQCD calculation.

The initial proper time $\tau_0$ is one of the parameters of the hydrodynamical
model: at the conditions of RHIC, the values $\tau_0=0.6$~fm and $\tau_0=1$~fm
have been chosen by the authors in the ideal \cite{kolb1} and viscous
\cite{rom2} cases, respectively. 
Since the value of $\tau_0$ has some impact on the final heavy
quark spectra (a lower value provides a longer propagation and exposes the
heavy quark to higher temperatures), we have also explored a scenario with a
very small value of $\tau_0$, by adjusting the maximal initial entropy density
$s_0$ (hence, through the EOS, the corresponding initial energy density) in such
a way to maintain the same particle density, using the relation
\begin{equation}
  \frac{dN}{dy}\sim s_0\tau_0.
\end{equation}
At the LHC regime, on the other hand, a full experimental constraint is not
available yet (although first data on particle density at mid-rapidity are
becoming available at $\sqrt{s}=2.76$~TeV \cite{Ali276}).
Here we have followed Refs.~\cite{Kes08,Cha08} (ideal) and
Ref.~\cite{rom3} (viscous), where the initial conditions have been fixed in
order to match, instead of the experimentally observed multiplicities, the
range of the predicted ones. Our set of initial conditions is summed up in
Table~\ref{tab:hydro}.

\begin{figure}
\includegraphics[clip,width=0.48\textwidth]{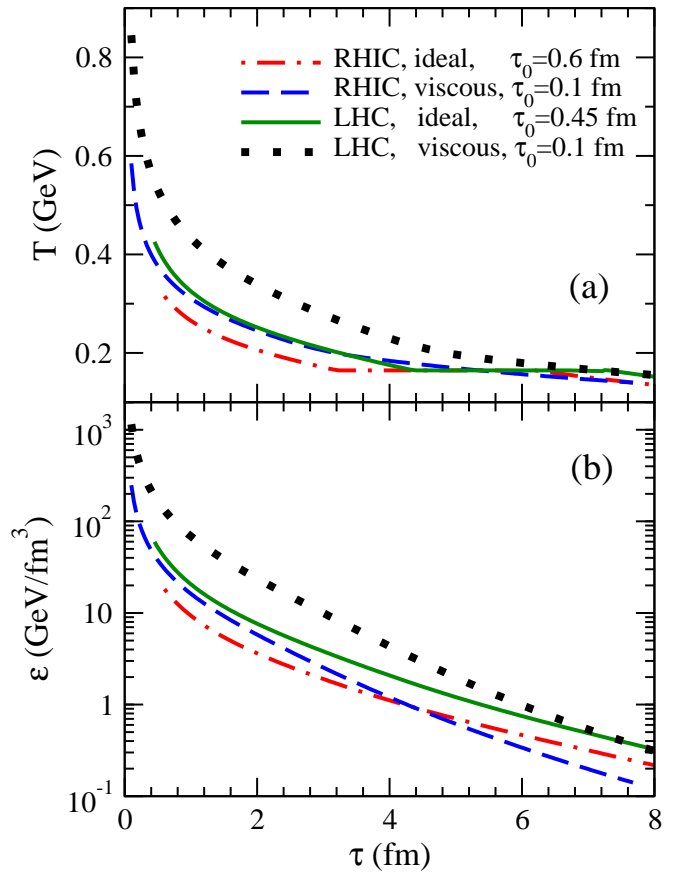}
\caption{(a) The temperature at the center of the fireball during
  hydrodynamical evolution for a few scenarios both at RHIC (impact parameter
  $b=8.44$~fm) and LHC ($b=8.77$~fm). (b) The energy density at the center of
  the fireball during hydrodynamical evolution for the same scenarios.}
\label{fig:hydro}
\end{figure}

As an example of the differences among the various hydrodynamical scenarios
that we have employed, in Fig.~\ref{fig:hydro} we display the temperature and the
energy density at the center of the fireball during the hydrodynamical evolution
for a few cases of semi-peripheral collisions. For every experimental setup, at
RHIC and LHC, we have reported the scenarios giving rise to the lowest and to
the highest temperatures attained during the evolution.

\subsection{Langevin dynamics in a relativistic fluid}

We wish to follow the propagation of $c$ and $b$ quarks --- initially produced 
in hard pQCD processes --- in the expanding, thermalized fireball of deconfined
matter that one expects to arise from a high energy heavy ion collision. 
The propagation of the heavy quarks in such a hot environment is modeled as a
Brownian motion by employing a relativistic Langevin equation
\cite{svet,tea,rapp,hira}: our implementation of the model has been discussed
in detail in a previous paper \cite{ber}, where we dealt with the simpler case
of a static homogeneous medium at rest. Here, we briefly summarize the essential
points of model.

The evolution with time of the momentum of a relativistic Brownian particle is
provided by the Langevin equation: 
\begin{equation}
  \frac{d\vec{p}}{dt}=-\eta_D(p)\vec{p}+\vec{\xi}(t),
\label{eq:lange_r_d}
\end{equation}
where the \emph{drag coefficient} $\eta_D(p)$ describes the deterministic
friction force acting on the heavy quark, whereas the term $\vec{\xi}$ accounts
for the random collisions with the constituents of the medium. 
The effect of the stochastic term is completely determined once its temporal
correlation function is fixed. The latter is usually assumed to be given by 
\begin{equation}
  \langle\xi^i(t)\xi^j(t')\rangle=b^{ij}(\vec{p})\delta(t-t'),
\label{eq:noise1}
\end{equation}
entailing that collisions at different time-steps are uncorrelated. The tensor
$b^{ij}(\vec{p})$ can be decomposed with a standard procedure according to
\begin{equation}
  b^{ij}(\vec{p})\equiv \kappa_L(p)\hat{p}^i\hat{p}^j+\kappa_T(p)
(\delta^{ij}-\hat{p}^i\hat{p}^j)
\label{eq:noise2}
\end{equation}
in terms of the coefficients $\kappa_{L/T}(p)$, which represent the squared
longitudinal/transverse momentum per unit time exchanged by the quark with the 
medium.
It is useful to introduce the related tensor
\begin{eqnarray}
  g^{ij}(\vec{p}) &\equiv&
    \sqrt{\kappa_L(p)}\hat{p}^i\hat{p}^j+\sqrt{\kappa_T(p)} 
    (\delta^{ij}-\hat{p}^i\hat{p}^j) \nonumber \\
  &\equiv& g_L(p)\hat{p}^i\hat{p}^j+g_T(p)(\delta^{ij}-\hat{p}^i\hat{p}^j),
\label{eq:gij}
\end{eqnarray}
which allows one to factor out the momentum dependence of the noise term in
Eq.~(\ref{eq:lange_r_d}), thus obtaining the equation
\begin{equation}
  \frac{dp^i}{dt} = -\eta_D(p)p^i + g^{ij}(\vec{p})\eta^j(t),
\label{eq:lange_r_d2}
\end{equation}
with
\begin{equation}
  \langle\eta^i(t)\eta^j(t')\rangle = \delta^{ij}\delta(t-t').
\label{eq:etaietaj}
\end{equation}
Finally, the drag coefficient $\eta_D(p)$ is fixed in order to ensure the
approach to equilibrium: for large times the momenta of an ensemble of heavy
quarks should be described by an equilibrium Maxwell-J\"uttner distribution.
Actually, the whole procedure depends on the discretization scheme
employed in the numerical solution of Eq.~(\ref{eq:lange_r_d2}), which belongs
to the class of the \emph{stochastic differential equations}
\cite{stocha,lau}. In the Ito discretization scheme one gets:
\begin{equation}
  \eta_D^{\rm Ito}(p)\equiv\frac{\kappa_L(p)}{2TE} 
    -\frac{1}{E^2}\left[(1\!-\!v^2)\frac{\partial\kappa_L(p)}{\partial v^2}
    +\frac{\kappa_L(p)-\kappa_T(p)}{v^2}\right], 
\label{eq:etaD}
\end{equation}
where $v$ is the quark velocity, $E$ its energy and $T$ the medium temperature.

The set of Eqs.~(\ref{eq:gij}-\ref{eq:etaD}) is defined in the rest frame of
the background medium and it allows one to study the quark propagation once the
transport coefficients are given. These depend on the medium temperature,
which in turn, in the expanding fireball, depends on the space-time position
occupied by the heavy quark.
Hence, to study the fate of a heavy quark through the quark-gluon plasma, we
adopt the following procedure:
\begin{itemize}
\item We determine the initial four-momentum $p^\mu$ and the initial space-time
  position $x^\mu$ of the heavy quark (in the laboratory system) as explained in
  Sect.~\ref{subsec:ini}.
\item Given the position $x^\mu$, we use the information from the hydrodynamic
  simulation (Sect.~\ref{subsec:hydro}) to obtain the fluid local temperature
  $T(x)$, velocity $u^\mu(x)$ and energy density $\varepsilon(x)$.
\item We check (see Sect.~\ref{subsec:hadronize}) whether the conditions for
  hadronization apply: in this case the procedure is ended; otherwise
\item we make a Lorentz transformation ($p^\mu \to \bar{p}^\mu$) to the fluid
  rest frame, employ Eqs.~(\ref{eq:gij}-\ref{eq:etaD}) to update the quark
  momentum ($\bar{p}^\mu \to \bar{p}'^\mu$) and boost it back to the laboratory
  ($\bar{p}'^\mu \to p'^\mu$).
\item We update the space-time step made by the quark in the rest frame
  ($\Delta \bar{x}^\mu=(\bar{p}^\mu/E_{\bar{p}})\Delta \bar{t}$), boost it to
  the laboratory ($\Delta \bar{x}^\mu \to \Delta x^\mu$) and use it to update
  the quark position ($x^\mu \to x'^\mu$).
\item Given the new momentum $p'^\mu$ and the new position $x'^\mu$ the
  procedure is started again until the conditions for hadronization are met.
\end{itemize}
The time step in the rest frame, which enters in updating the quark position and
also the quark momentum through the Langevin equation, in our calculations has
the value $\Delta \bar{t}=0.02$~fm.

\subsection{Heavy-quark transport coefficients}

\begin{figure*}
\begin{center}
\includegraphics[clip,width=0.9\textwidth]{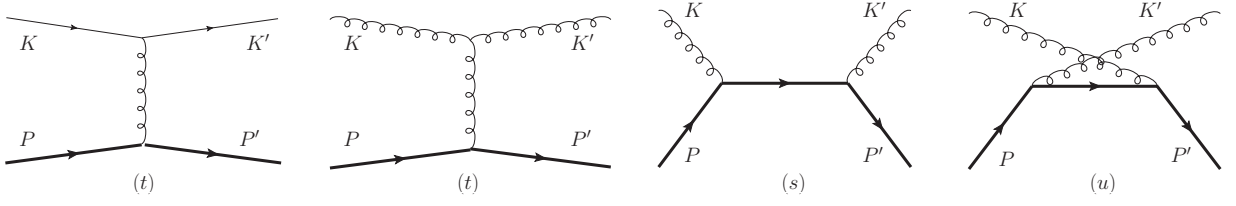}
\caption{The tree level diagrams for the hard scattering of a heavy quark off
  a light (anti-)quark of the thermal bath and for the hard scattering (in the
  $t$, $s$ and $u$ channels) of a heavy quark off a gluon.}  
\label{fig:quarkgluon}
\end{center}
\end{figure*}

In the approach to stochastic dynamics based on the (relativistic) Langevin
equation, the information on the properties of the medium crossed by a Brownian
particle is encoded in the drag and momentum-diffusion coefficients $\eta_D$ and $\kappa_{T/L}$. Following the random motion of the heavy quarks in the QGP
requires then the calculation of the above coefficients, starting from the
microscopic theory, i.e. finite temperature QCD. 
In accord with Refs.~\cite{pei1,pei2}, we accomplish this task by separating soft
and hard collisions depending upon the value of the Mandelstam variable $t\equiv
\omega^2-q^2$ that characterizes the elementary scattering, and summing at the
end the two contributions\footnote{Note that this differs from what is usually 
done in the literature (see, e.g., Ref.~\cite{tho12}), where the separation
between hard and soft scatterings is related to the spatial momentum $q$
exchanged in the collision.}.
 
Soft scatterings --- corresponding to $\sqrt{|t|}\sim gT$ --- occur quite
frequently, with a typical mean free path $\lambda_{\rm mfp}^{\rm soft}\sim
1/g^2T$. Being related to the exchange of long-wavelength gluons, they require
a careful treatment of medium effects, which can be achieved by describing
these soft collisions within the HTL approximation. 
Hard scatterings --- corresponding to $\sqrt{|t|}\gsim T$ --- are more rare,
the mean free path being in this case $\lambda_{\rm mfp}^{\rm hard}\sim
1/g^4T$, but, since they cause a sizable momentum exchange, their contribution
to the transport coefficients is substantial and will be evaluated, as already
mentioned, within a microscopic kinetic calculation based on pQCD. 
We set $|t|^*\sim m_D^2$, $m_D$ being the Debye mass, as the intermediate
cutoff. Clearly, the final results should not be too much affected by the choice 
of $|t|^*$ and we have verified that this is actually the case, even for
temperatures of experimental interest, where the coupling is not really small. 

The coupling constant $g$ has been chosen to run according to the two-loop QCD 
beta-function, with $\Lambda_{\rm QCD}=261$~MeV, as in Ref.~\cite{za}.
The coupling constant displays a strong dependence on the unknown value of the
temperature dependent scale $\mu\propto T$ and this has important consequences
on the final observables, as we shall see in the following.
To shed light on this important issue, we shall explore the effect on the
calculated quantities for a wide range of values, $\pi T\le\mu\le2\pi T$.

\subsubsection{Soft collisions: hard thermal loops}

The contribution of soft collisions to the transport coefficients has already
been obtained in our previous work \cite{ber}, to which we refer for details.
However, due to our choice of setting $|t|^*$ as an intermediate cutoff, it is
necessary to rewrite the formulas in Ref.~\cite{ber} for $\kappa_{T/L}$  
employing $|t|\equiv q^2-\omega^2$ and $x\equiv \omega/q$ as integration
variables.
One gets:
\begin{eqnarray}
  \kappa_T^{\rm soft} &=& \frac{C_F g^2}{8\pi^2 v}
    \int_0^{|t|^*} d|t| \int_{0}^{v} dx \frac{|t|^{3/2}}{2(1-x^2)^{5/2}}
    \overline{\rho}(|t|,x) \nonumber \\
  && \quad \times \left(1-\frac{x^2}{v^2}\right) \coth\left(
    \frac{x\sqrt{|t|/(1-x^2)}}{2T}\right) 
\end{eqnarray}
and
\begin{eqnarray}
  \kappa_L^{\rm soft} &=& \frac{C_F g^2}{4\pi^2 v} 
    \int_0^{|t|^*}d|t| \int_{0}^{v} dx \frac{|t|^{3/2}}{2(1-x^2)^{5/2}}
    \overline{\rho}(|t|,x) \nonumber \\
  && \quad \times \frac{x^2}{v^2} \coth\left(
    \frac{x\sqrt{|t|/(1-x^2)}}{2T}\right),
\end{eqnarray}
for the transverse and longitudinal momentum diffusion coefficients,
respectively. In the above, $v$ is the quark velocity, $C_F=4/3$ the quark 
Casimir factor and 
\begin{equation}
  \overline{\rho}(|t|,x)\equiv \rho_L(|t|,x)+(v^2-x^2)\rho_T(|t|,x),
\end{equation}
where $\rho_{L/T}$ are the continuum parts of the HTL gluon spectral functions:
\begin{eqnarray}
  &&\rho_L(|t|,x) = \pi m_D^2 x \left\{\left[\frac{|t|}{1-x^2} \right.\right.
    \nonumber \\
  && \quad \left.\left. + m_D^2\left(1-\frac{x}{2}\ln\left|
    \frac{x+1}{x-1}\right|\right)\right]^2 
    + \left[\pi m_D^2\frac{x}{2}\right]^2 \right\}^{-1} \nonumber \\  
  &&\rho_T(|t|,x) = \pi m_D^2\frac{x}{2}(1-x^2) \left\{\left[|t| 
    + m_D^2\frac{x}{2}\phantom{\frac{x}{x}^2}  \right.\right. \\
  && \left.\left. \times \left(1+\frac{1-x^2}{2x}\ln\left|
    \frac{x+1}{x-1}\right|\right)\right]^2 +\left[\frac{\pi}{2}
    m_D^2\frac{x}{2}(1-x^2)\right]^2 \right\}^{-1}. \nonumber 
\end{eqnarray}

\subsubsection{Hard collisions: perturbative QCD}

\begin{figure}[!b]
\begin{center}
\includegraphics[clip,width=0.48\textwidth]{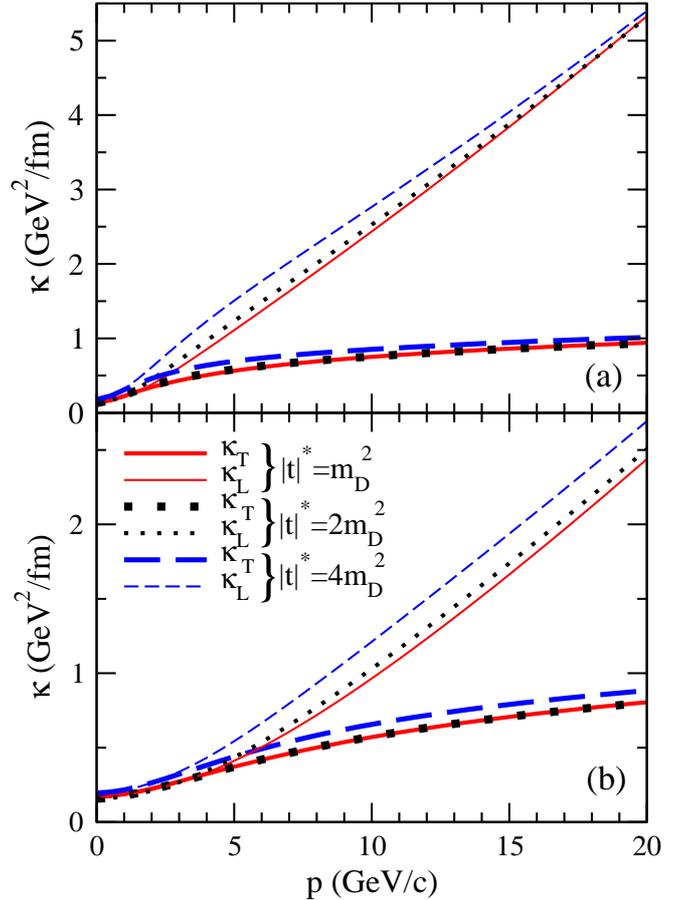}
\caption{The charm (upper panel) and bottom (lower panel) momentum diffusion
  coefficients $\kappa_{T/L}(p)$ resulting from the sum of the soft and hard
  contributions. The sensitivity to the intermediate cutoff $|t|^*\sim m_D^2$
  is very mild. The curves refer to the temperature $T=400$~MeV, with the
  coupling $g$ evaluated at the scale $\mu=(3/2)\pi T$.} 
\label{fig:cutoff}
\end{center}
\end{figure}

The contribution from hard collisions to the transport coefficients
$\kappa_{T/L}$ can be calculated starting from their microscopic definition,
namely
\begin{equation}
  \kappa_T\equiv\frac{1}{2}\left\langle\frac{\Delta q_T^2}{\Delta
  t}\right\rangle \quad {\rm and} \quad \kappa_L\equiv\left\langle\frac{\Delta
  q_L^2}{\Delta t}\right\rangle, 
\end{equation}
and by weighting the interaction rate with the squared transverse and
longitudinal momentum exchanged in the collisions with gluons and (anti-)quarks
of the medium. One has
\begin{equation}
  \kappa_{T/L}^{\rm hard}=\kappa_{T/L}^g+\kappa_{T/L}^q,
\end{equation}
where (employing the notation $\int_k\equiv \int d\vec{k}/(2\pi)^3$)
\begin{eqnarray}
  &&\kappa_T^{g/q} = \frac{1}{2E}\int_k\frac{n_{B/F}(k)}{2k}
    \int_{k'}\frac{1\pm n_{B/F}(k')}{2k'}\int_{p'}\frac{\theta(|t|-|t|^*)}{2E'}
    \nonumber \\ 
  &&\quad\times(2\pi)^4\delta^{(4)}(P+K-P'-K')\left|\overline{{\cal
    M}}_{g/q}(s,t)\right|^2 \frac{q_T^2}{2}
\label{eq:kthard} 
\end{eqnarray}
and
\begin{eqnarray}
  &&\kappa_L^{g/q} = \frac{1}{2E}\int_k\frac{n_{B/F}(k)}{2k} \int_{k'}
    \frac{1\pm n_{B/F}(k')}{2k'}\int_{p'}\frac{\theta(|t|-|t|^*)}{2E'}
    \nonumber \\ 
  && \quad \times(2\pi)^4\delta^{(4)}(P+K-P'-K')\left|\overline{{\cal
    M}}_{g/q}(s,t)\right|^2 q_L^2.
\label{eq:klhard} 
\end{eqnarray}
In the above expressions, $s=(P+K)^2$, $Q\equiv(\omega,\vec{q})\equiv P-P'$, 
$q_T/q_L$ are the transverse/longitudinal components of $\vec{q}$ and 
$n_{B/F}$ are the Bose/Fermi distributions; the squared amplitudes are averaged
over the internal degrees of freedom (colour and spin) of the incoming heavy
quark and summed over the internal degrees of freedom of the initial and final 
state partons in the medium: $N_c^2-1$ colours and two transverse 
polarizations for a gluon, $N_c$ colours and two helicities for a light quark. 
Furthermore a factor $2N_f$ is assigned to 
$\left|\overline{{\cal M}}_q\right|^2$, accounting for the identical
contribution provided by all the flavours of light quarks and anti-quarks in the
medium. The squared amplitudes $\left|\overline{{\cal M}}_{g/q}\right|^2$
defined in such a way can be obtained starting from the results given in
Ref.~\cite{com} and their expressions are reported in Appendix~\ref{app:hard}. 
The quark contribution is then obtained by squaring the corresponding amplitude
in Fig.~\ref{fig:quarkgluon}, given by a simple gluon exchange in the
$t$-channel. The gluon contribution is more cumbersome and requires squaring
the sum of the $t$, $s$ and $u$-channel diagrams in Fig.~\ref{fig:quarkgluon}.  

\begin{figure}
\begin{center}
\includegraphics[clip,width=0.48\textwidth]{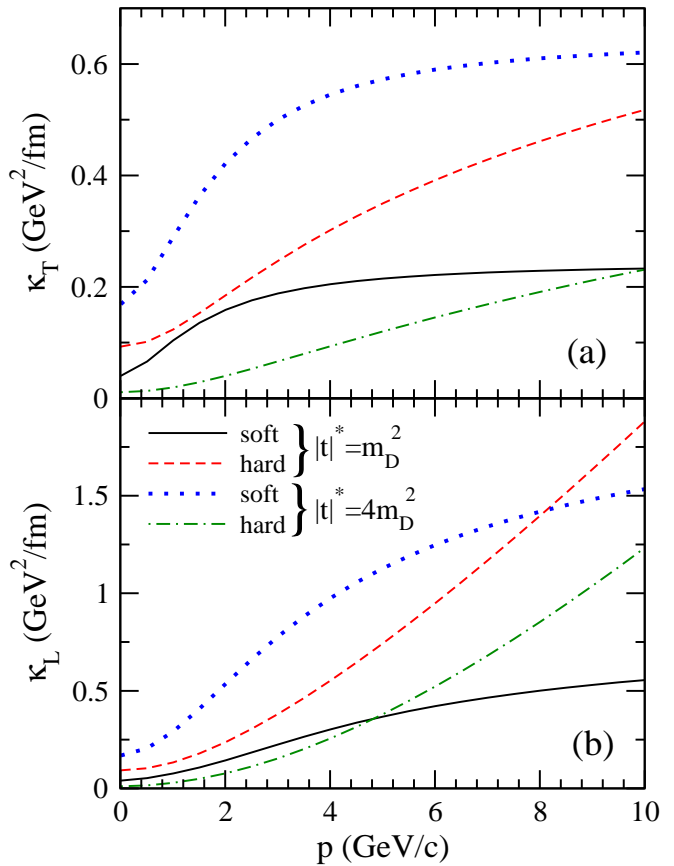}
\caption{The separate soft and hard contributions to $\kappa_T(p)$ (upper panel)
  and $\kappa_L(p)$ (lower panel) for different choices of the intermediate
  cutoff $|t|^*\sim m_D^2$. The curves refer to the case of a $c$ quark, in a
  medium at $T=400$ MeV, with the coupling $g$ evaluated at the scale
  $\mu=(3/2)\pi T$.} 
\label{fig:hardsoft}
\end{center}
\end{figure}

\begin{figure}
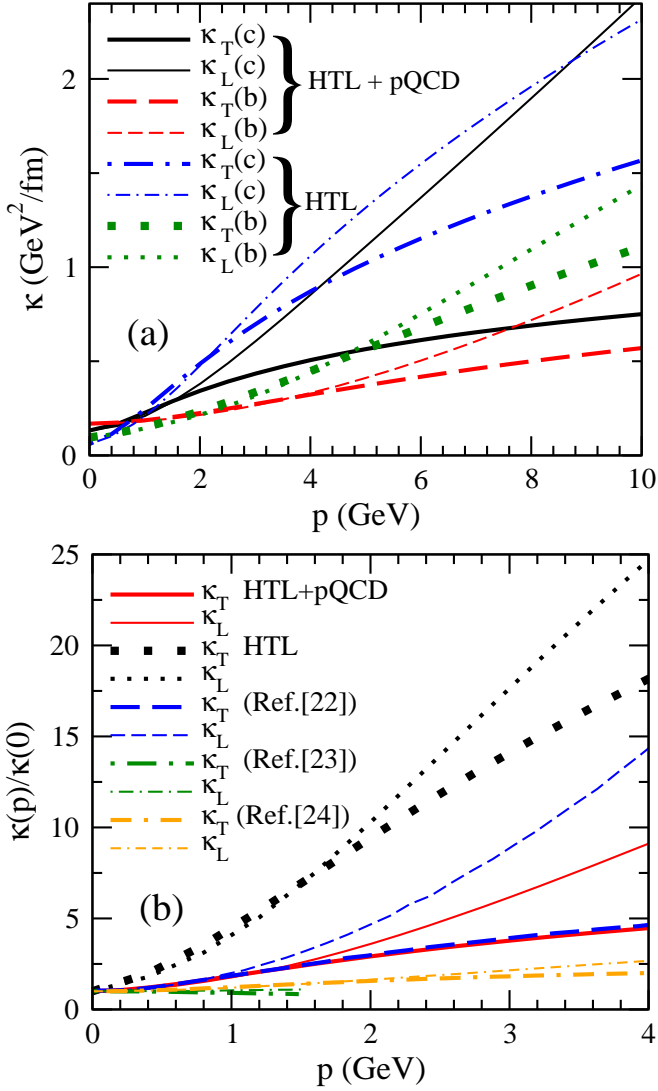

\begin{center}
\includegraphics[clip,width=0.48\textwidth]{fig_transport_cb_fullvsHTL.eps}
\includegraphics[clip,width=0.48\textwidth]{fig_kappa_resc_compare.eps} 
\caption{(a) Transport coefficients, for both $c$ and $b$ quarks
  at $T=400$~MeV and $\mu=(3/2)\pi T$, arising from the separate treatment of
  soft (HTL) and hard (pQCD) collisions (with $|t^*|=m_D^2$) compared with the
  pure HTL results of Ref.~\protect\cite{ber}.
  (b) Comparison of our results for $c$ quarks (with $|t^*|=m_D^2$
  and $\mu=(3/2)\pi T$) to the outcomes of several models
  \protect\cite{ber,tea,rapp,aic} for the transport coefficients
  $\kappa_{T/L}(p)$ at $T=200$~MeV.}
\label{fig:comparison}
\end{center}
\end{figure}
\subsubsection{Charm and bottom transport coefficients}

In this section we display the outcome of our calculation of the heavy quark
transport coefficients in the QGP and compare our results with previous
findings for the same coefficients available in the literature 
\cite{ber,tea,rapp,aic}.

As we have already mentioned, following the heavy-quark stochastic propagation
in the hot environment produced in heavy ion collisions requires the knowledge
of the corresponding transport coefficients for a range of temperatures from
$T\sim T_c$ (where actually we expect our weak-coupling approach to
underestimate their value) up to $T\sim 1$ GeV  (for the case of the LHC 
scenarios with the highest initial energy-density).

In Fig.~\ref{fig:cutoff} we show our findings for $\kappa_{T/L}(p)$ for $c$ and
$b$ quarks, at the temperature $T=400$ MeV; the coupling $g$ has been 
evaluated at the scale $\mu=(3/2)\pi T$. 
Two important features are apparent in Fig.~\ref{fig:cutoff}: the larger growth
with the momentum $p$ of $\kappa_L(p)$ with respect to $\kappa_T(p)$ and the
very mild sensitivity to the value of the intermediate cutoff $|t|^*\sim
m_D^2$. This last finding occurs in spite of the fact that, at the
experimentally accessible temperatures, the coupling is not small. Such an
occurrence supports the validity of the adopted approach. 

A deeper insight can be gained by looking at Fig.~\ref{fig:hardsoft}, where we
plot separately the soft and hard contributions to $\kappa_T(p)$ and
$\kappa_L(p)$ for the case of $c$ quarks. 

We finally compare our results for the heavy-quark transport coefficients to
the ones obtained using other models. 
We start with our previous work \cite{ber}, where the HTL approximation had
been applied to any value of the momentum exchange, up to a cut-off 
$q_{\rm max}$. Results for $c$ and $b$ quarks are shown in 
Fig.~\ref{fig:comparison}a and compared with the present, more
realistic, treatment. The major difference one observes is in the growth of
$\kappa_T(p)$ with $p$, which is faster in the pure HTL approach. 

In Fig.~\ref{fig:comparison}b we summarize the outcomes of
several models which can be found in the literature: the present approach
(pQCD+HTL), the pure HTL result~\cite{ber} and the findings of
Refs.~\cite{tea,rapp,aic}. 
In the approach of Ref.~\cite{tea}, the coefficients are obtained from a 
kinetic pQCD calculation, with the value of $\kappa(p=0)$ tuned by
hand in order to explore both strongly and weakly-coupled scenarios. 
In Ref.~\cite{rapp} the mechanism responsible for the heavy-quark
thermalization is resonant scattering, with the temporary formation of 
finite-width D-mesons in the plasma. The authors of Ref.~\cite{aic}, on the
other hand, resort to an effective running coupling constant and to a value of
the Debye mass lower than the thermal QCD prediction.
Notice how all the models based on pQCD tend to favor small-angle scattering.

\subsection{Hadronization and decay}
\label{subsec:hadronize}

\begin{figure}
\begin{center}
\includegraphics[clip,width=0.48\textwidth]{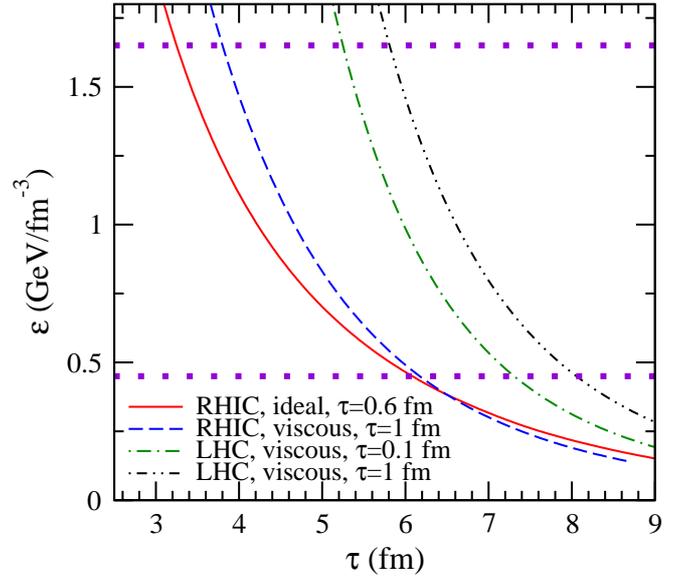}
\caption{Energy density in the center of the fireball at proper times
  around the deconfinement transition for a few hydrodynamical setups; the
  dotted lines correspond to $\varepsilon_{\rm QGP}=1.65$~GeV/fm$^3$ and 
  $\varepsilon_{\rm H}=0.45$~GeV/fm$^3$.}
\label{fig:phasetrans}
\end{center}
\end{figure}

The conversion of the heavy quarks into hadrons requires two distinct steps: 
first, one has to decide when the quark is going to hadronize and then to apply
a specific model for the transition.
In the ideal hydrodynamics model of Refs.~\cite{kolb1,kolb2,kolb3} a first
order phase transition occurs at the temperature $T_c=164$~MeV, lasting for a 
couple of fm and during which the energy density drops from 
$\varepsilon_{\rm QGP}=1.65$~GeV/fm$^3$ to $\varepsilon_{\rm H}=0.45$~GeV/fm$^3$.
In the viscous hydrodynamics model of Refs.~\cite{rom1,rom2,rom3}, on the other
hand, a more realistic crossover, slightly faster because of viscosity, is
employed. 
In Fig.~\ref{fig:phasetrans} we display the energy density in the center of the
fireball around the QGP/hadron-matter transition for a few hydrodynamical
scenarios. The dotted lines mark the position of $\varepsilon_{\rm QGP}$ and 
$\varepsilon_{\rm H}$: we assume that hadronization is going to take place in
this region of mixed phase. 

Introducing the fraction of QGP in the mixed phase \cite{hira} according to
\begin{equation}
  f_{\rm QGP} = 
    \frac{\varepsilon-\varepsilon_{\rm H}}
    {\varepsilon_{\rm QGP}-\varepsilon_{\rm H}},
\end{equation}
we stop the Langevin propagation of the heavy quark according to the following
prescription:  
\begin{itemize}
\item we extract the medium energy density at the heavy-quark space-time
  position;
\item if $f_{\rm QGP}$ is larger than one, the Langevin propagation is carried
  on another step; otherwise
\item we treat $1-f_{\rm QGP}$ as a transition probability: given $f^i_{\rm
  QGP}$ --- the fraction of QGP at the $i$-th propagation step --- we draw a
  random number $h$; if $h\ge f^i_{\rm QGP}/f^{i-1}_{\rm QGP}$ then
  hadronization has occurred, otherwise a new Langevin step is taken and the
  procedure repeated at the new position. 
\end{itemize}
In this way the transition from quarks to hadrons is made occur over the
whole mixed phase.

\subsubsection{Hadronization}

\begin{figure*}
\begin{center}
\includegraphics[clip,width=0.85\textwidth]{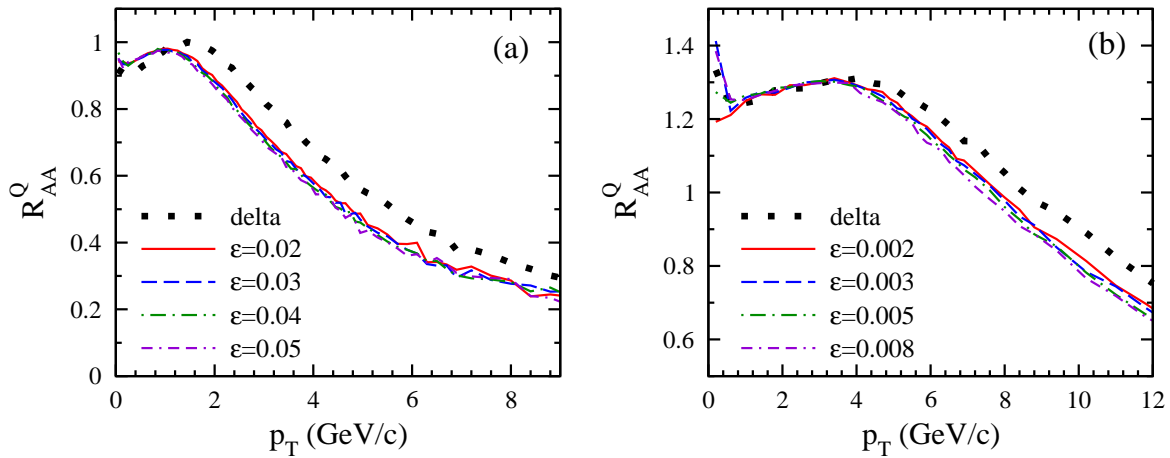}
\caption{Nuclear modification factor $R_{AA}$ vs. $p_T$ for open charm (left)
  and open beauty (right) hadrons for different values of the $\varepsilon$
  parameter in the Peterson fragmentation function; the dotted lines represent
  the delta function limit in the fragmentation function, $D(z)=\delta(1-z)$; 
  in these examples we
  have used ideal hydrodynamics at RHIC for $b=8.44$~fm and $\mu=2\pi T$.} 
\label{fig:epsiloneffect}
\end{center}
\end{figure*}

Two mechanisms of hadronization should in principle be implemented, namely
recombination of the heavy quark with a light quark from the medium and
fragmentation. The coalescence approach has been shown to be important in the
low momentum part of the spectra at RHIC \cite{rapp}, say for $p_T\lsim3$~GeV/c.
In this work, we limit ourselves to let the heavy quarks hadronize by
fragmentation, hence our results should be viewed as realistic for relatively
large momenta, $p_T\gsim3$~GeV/c at RHIC energy. Note, however, that the region
where recombination is dominant might change at LHC energies. 

Open charm and open beauty hadrons are produced from the charm and bottom 
quarks via a two-step Monte Carlo procedure.
First, the resulting hadron species is assigned according to measured 
fragmentation ratios. 
In the case of charm quarks, four hadronic final states have been considered, 
namely $\rm D^0$, $\rm D^+$, $\rm D_s^+$ and $\Lambda_c^+$, together with their 
corresponding anti-particles.
The fragmentation ratios are taken from Table 4 of Ref.~\cite{ZEUS}.
For bottom quarks, the considered hadronic states have been $\rm B^0$, $\rm
B^+$, $\rm B_s^0$ and $\Lambda_b^0$, together with their corresponding
anti-particles. 
The fragmentation ratios have been taken from~\cite{HFAG} with the
only modification of assigning to the $\Lambda_b^0$ final state the
probability for a bottom quark to fragment into b-baryons.

The second step is the Monte Carlo generation of the hadron momenta.
This has been done starting from the momentum components ($p_x$, $p_y$ and
$p_z$) of the heavy quark out-coming from the Langevin evolution and sampling,
from a Peterson fragmentation function~\cite{PETERSON}, the fraction $z$ of 
quark momentum that is taken by the produced hadron:
\begin{equation}
  D(z)=\frac{k}{z} \cdot \left( 1-\frac{1}{z}-\frac{\varepsilon}{1-z}
    \right)^{-2}.
\end{equation}
The parameter $k$ has been set in order to have the fragmentation function
normalized to 1, while for $\varepsilon$ the values 0.04 and 0.005 have been
used for charm and bottom quarks, respectively.
They have been chosen by comparing the fragmentation function shapes with the
ones computed in Ref.~\cite{Braaten} on a pQCD or Heavy Quark Effective Theory
basis. 
To check the effect of the choice of the $\varepsilon$ parameter on the
resulting nuclear modification factor $R_{AA}$ (see Eq.~\ref{eq:RAA}), the whole
fragmentation procedure has been repeated four times varying the value of $\varepsilon$ in
the range yielding a reasonable agreement with the calculation of
Ref.~\cite{Braaten}. 
The obtained $R_{AA}$ for open charm and open beauty hadrons as a function of 
the transverse momentum $p_T$ for different choices of $\varepsilon$ is shown
in Fig.~\ref{fig:epsiloneffect} for a specific case of $\sqrt{s}$, centrality
and hydrodynamical scenario.
We have checked that variations of the parameter $\varepsilon$ by a factor two
results generally in a $\lsim3\%$ effect on the $R_{AA}$ of open heavy-flavoured
hadrons.

\subsubsection{Decays of heavy-flavour hadrons into electrons}

In order to compare our calculations to the experimental data for the 
non-photonic single-electron ($e^{\pm}$) transverse spectra and for
the corresponding nuclear modification factor $R_{AA}$ and elliptic flow
coefficient $v_2$, each charm or bottom hadron originated by the fragmentation
of a heavy quark (as described in the previous section), is forced to decay
into electrons.

This task is performed by using the decay routines implemented inside the
PYTHIA event generator \cite{PYTHIA}. However, a revised table of the branching
ratios for the dominant decay channels of each heavy-quark hadron has been
employed, in order to replace the default values which are by now obsolete.
Such table has been drawn up by taking the measured values of the
branching ratios, as reported on the latest Particle Data Group review
\cite{PDG}.

Only hadronic and semi-leptonic decay channels of charm and beauty hadrons with
branching ratio $\Gamma > 10^{-4}$ have been included. Therefore the
contribution of each hadron to the final electron spectra has been properly
weighted by the corresponding total branching ratio.

In this way, all contributions to the electron yield (i.~e., prompt electrons
from $B \rightarrow e$ or $D \rightarrow e$ decays, and secondary electrons from 
a cascade process $B \rightarrow D \rightarrow e$) are naturally implemented 
in our simulation scheme, as well as the additional ``background'' contribution 
from the decay chain $B \rightarrow J/\psi \rightarrow e^{+} e^{-}$. Indeed, 
the latter contribution has been subtracted from the total measured electron 
yields only in the most recent analyses of the PHENIX data \cite{phenix2}.

The electron spectra originated from charm ($e_c$) and bottom ($e_b$) decays
are then combined into a unique spectrum ($e_{c+b}$) with an appropriate
weight, which accounts for the corresponding production cross section of the
parent heavy quark.

\begin{figure*}
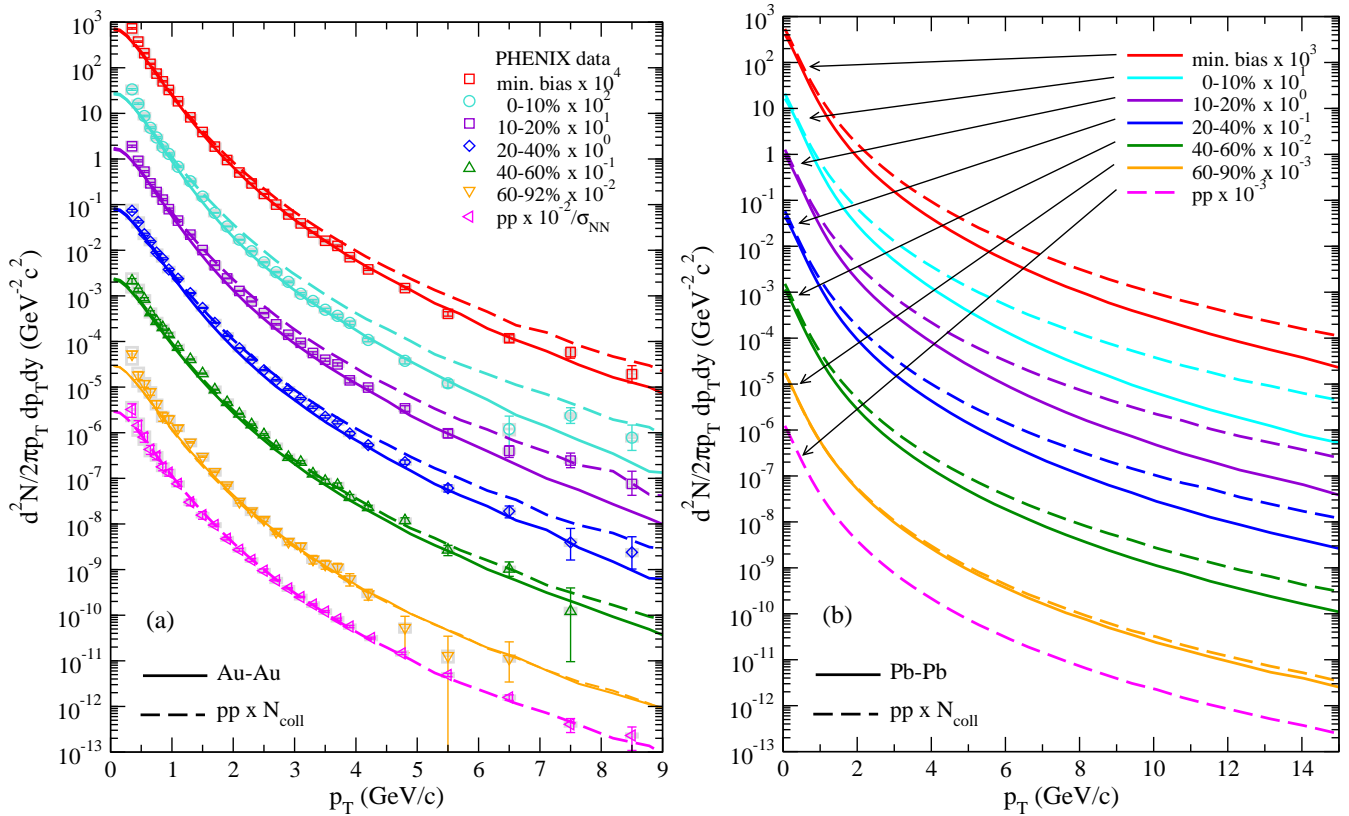

\begin{center}
\includegraphics[clip,width=0.485\textwidth]{fig_idcs_AA_RHIC.eps}
\includegraphics[clip,width=0.485\textwidth]{fig_idcs_AA_LHC5.eps}
\caption{(a) Invariant yields of electrons [$(e^++e^-)/2$], from heavy-flavour 
  decay in Au-Au collisions at $\sqrt{s}=200$~GeV, at mid-rapidity as a
  function of transverse momentum (scaled for clarity).
  The experimental points represent PHENIX
  data~\protect\cite{phenix,phenix2}; both statistical (error bars) and
  systematic (gray boxes) errors are displayed. 
  The dashed curves represent the contributions coming from electrons
  originating from charm and bottom quarks generated by POWHEG with inclusion of
  transverse broadening and rescaled by $N_{\rm coll}$. The solid curves are
  our results after propagation in the fireball (viscous hydrodynamics with
  $\tau_0=1$~fm, $\mu=3\pi T/2$).
  (b) Invariant yields of electrons [$(e^++e^-)/2$], from heavy-flavour 
  decay in Pb-Pb collisions at $\sqrt{s}=5.5$~TeV, at mid-rapidity as a
  function of transverse momentum (scaled for clarity). Dashed and solid lines
  as in panel (a). 
}
\label{fig:idcs_AA}
\end{center}
\end{figure*}

\section{Heavy-flavour spectra in nucleus-nucleus collisions}
\label{sec:res}

\begin{table}[b]
\caption{\label{tab:CC} The centrality classes and the corresponding average
  impact parameters at the kinematics of RHIC and LHC.
}
\begin{center}
\begin{tabular}{|rr|rr|}
\hline
    \multicolumn{2}{|c|}{Au-Au ($\sqrt{s}=200$ GeV)} & 
    \multicolumn{2}{c|}{Pb-Pb ($\sqrt{s}=5.5$ TeV)}    \\
\hline
   $C_1$-$C_2$  & $b$ (fm) & 
   $C_1$-$C_2$  & $b$ (fm) \\
\hline
  0-10\%  &  3.27 &  0-10\% &  3.45 \\
 10-20\%  &  5.78 & 10-20\% &  6.11 \\
 20-40\%  &  8.12 & 20-40\% &  8.58 \\
 40-60\%  & 10.51 & 40-60\% & 11.11 \\
 60-92\%  & 12.80 & 60-90\% & 13.45 \\
  0-92\%  &  8.44 &  0-90\% &  8.77 \\
\hline
\end{tabular}
\end{center}
\end{table}

We have considered two kinematic regimes for high energy heavy ion collisions:
Au-Au at $\sqrt{s}=200$~GeV as in the PHENIX experiment at RHIC
\cite{phenix,phenix2} and Pb-Pb at $\sqrt{s}=5.5$~TeV, the highest energy that
should be attained with $AA$ collisions at the LHC.

In the PHENIX experiment the heavy-flavour data have been presented in several
centrality classes $C_1$-$C_2$, corresponding to the fraction of the geometric
cross section with impact parameter in the range $b_1<b<b_2$:
\begin{equation}
  f_{C_1{\rm -}C2}=\frac{\int_{b_1}^{b_2}db\,b[1-\exp(\sigma_{\rm NN}T_{AB}(b)]}
    {\int_0^\infty db\,b[1-\exp(\sigma_{\rm NN}T_{AB}(b)]},
\end{equation}
where $T_{AB}(b)=\int d\vec{s} T_A(\vec{s}+\vec{b}/2) T_B(\vec{s}-\vec{b}/2)$ is
the nuclear overlap function and the nuclear profile function in the transverse
plane $T_A(\vec{s})$ is given in Eq.~(\ref{eq:TA}). Our calculations are
performed for a given impact 
parameter $b$, so for every centrality class we have to evaluate an average
impact parameter. This can be done by equating the number of collisions in a
given centrality class $<N_{\rm coll}>_{C_1{\rm -}C_2}=
\sigma_{\rm NN}<T_{AB}>_{C_1{\rm -}C_2}=
\sigma_{\rm NN}\int_{b_1}^{b_2}db\,b T_{AB}(b)/\int_{b_1}^{b_2}db\,b$ to the
number of collisions at a given impact parameter 
$N_{\rm coll}(b)=\sigma_{\rm NN} T_{AB}(b)$ \cite{dEn03}. 
In Table~\ref{tab:CC} we show the centrality classes and the corresponding
impact parameters that we have considered in the present work.  At the LHC
energy, lacking the experimental data, we have kept a partition similar to the
one employed of RHIC.

\subsection{Differential spectra}

In Fig.~\ref{fig:idcs_AA} one can see the invariant yields of electrons from
heavy-flavour decays in Au-Au and Pb-Pb collisions, for a specific choice of
the hydrodynamical scenario (viscous hydrodynamics with $\tau_0=1$~fm) and of 
the QCD scale factor ($\mu=3\pi T/2$). Binary-scaled $pp$ spectra are also
shown as dashed lines. We are aware that cross sections are not the best
observable for studying the effect of the nuclear medium on the propagation of
the heavy quarks; for this purpose the nuclear modification factor is much more
suitable. It is however worth noting how the formalism is able to give a fair
description of the PHENIX data over many orders of magnitude. Of course, this is 
also a consequence of the good performance of the pQCD outcome for the heavy
quark production in NN collisions, as already discussed in
Sect~\ref{subsec:ini}.

\subsection{Nuclear modification factor}
\label{subsec:RAA}

\begin{figure*}
\begin{center}
\includegraphics[clip,width=0.85\textwidth]{fig_RAA_mu.eps}
\caption{(a) The heavy-quark nuclear modification factor at RHIC 
  for viscous hydrodynamics ($\tau_0=1$~fm), $b=8.44$~fm and different choices
  of the QCD scale factor; both $c$ (light lines) and $b$ (heavy lines) quarks
  are shown. (b) As in panel (a), but for LHC at $b=8.77$~fm.
}
\label{fig:RAA_mu}
\vskip 0.2cm
\includegraphics[clip,width=0.85\textwidth]{fig_RAA_hydro.eps}
\caption{(a) The heavy-quark nuclear modification factor at RHIC 
  for $\mu=3\pi T/2$, $b=8.44$~fm and different hydrodynamical scenarios; both
  $c$ (light lines) and $b$ (heavy lines) quarks are shown. (b) As in panel
  (a), but for LHC at $b=8.77$~fm.
}
\label{fig:RAA_hydro}
\vskip 0.2cm
\includegraphics[clip,width=0.85\textwidth]{fig_RAA_QAe.eps}
\caption{(a) The nuclear modification factor of heavy quarks,
  open heavy-flavour hadrons and electrons at RHIC for $\mu=3\pi T/2$, viscous
  hydrodynamics ($\tau_0=1$~fm) and $b=8.44$~fm; both the charm (light lines)
  and bottom (heavy lines) sectors are shown. (b) As in panel (a), but for LHC
  at $b=8.77$~fm.
}
\label{fig:RAA_QAe}
\end{center}
\end{figure*}

\begin{figure*}
\begin{center}
\includegraphics[clip,width=0.8\textwidth]{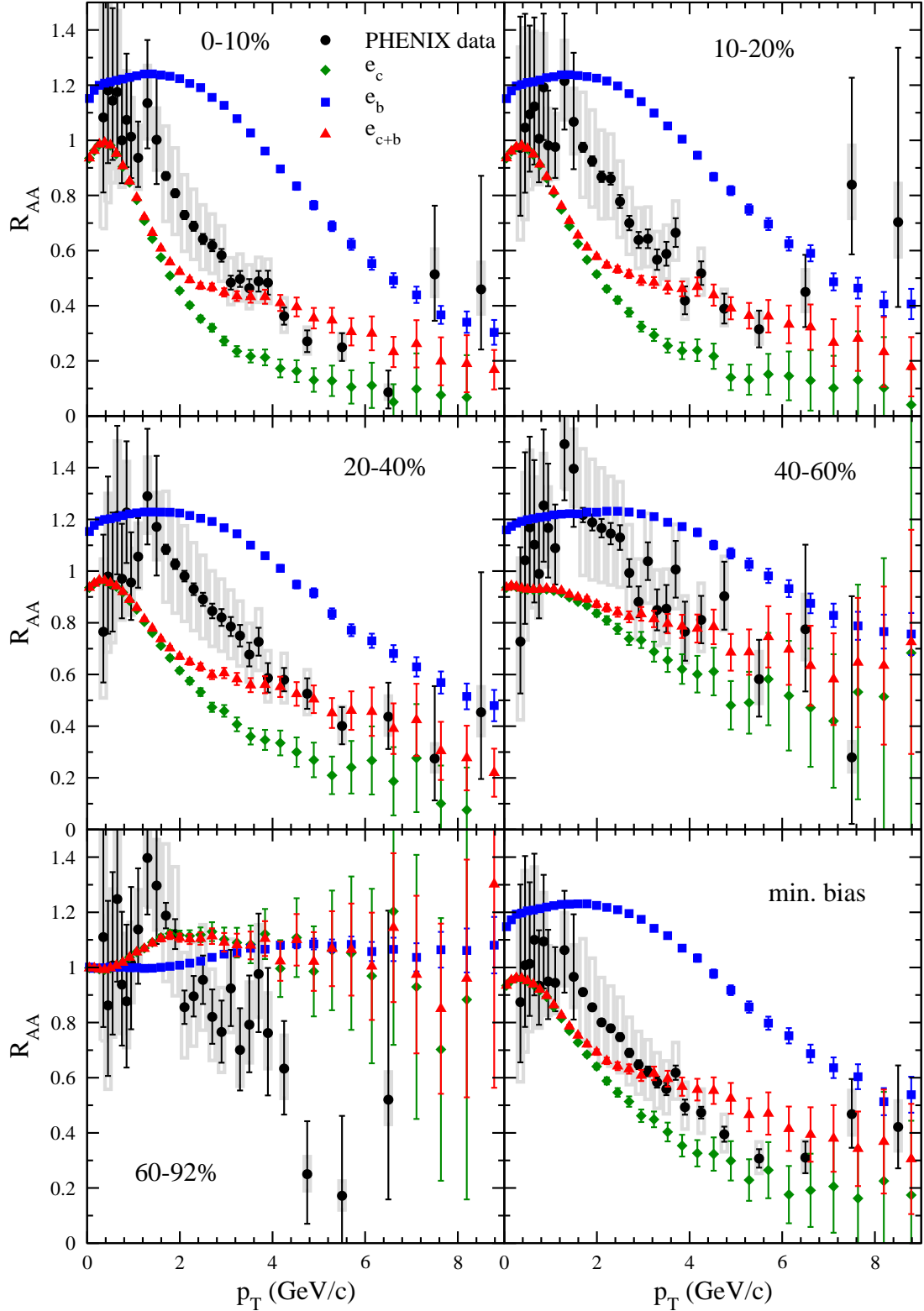}
\caption{The nuclear modification factor of open heavy-flavour electrons at
  RHIC for 
  $\mu=3\pi T/2$ and viscous hydrodynamics ($\tau_0=1$~fm) in various
  centrality classes. The circles are data from the PHENIX experiment
  \protect\cite{phenix,phenix2}, including both statistical (error bars) and
  systematic (grey boxes) errors; the other points (with the statistical errors
  shown) are the outcome of our calculations for electrons originating from
  charm and bottom quarks and their combination.
}
\label{fig:RAA_phenix}
\end{center}
\end{figure*}

Let us consider now the nuclear modification factor,
\begin{equation}
  R_{AA}(p_T) = \frac{dN_{AA}/dp_T}{N_{\rm coll}dN_{pp}/dp_T}.
\label{eq:RAA}
\end{equation}
In order to explore the effect and the relative importance of those ingredients
of the model that are not fully under control, that is the QCD scale factor and
the hydrodynamical scenario, it is instructive to start by studying the nuclear
modification factor at quark level ($R^{Q}_{AA}$) first, which is the one 
obtained from Eq.~(\ref{eq:RAA}) by using the spectra before hadronization
(keeping the same pseudorapidity window as for the electrons).
Thus, in Fig.~\ref{fig:RAA_mu} we display the charm and bottom nuclear
modification factors, for the case of viscous hydrodynamics ($\tau_0=1$~fm) and
minimum bias collisions, at both RHIC and LHC energies. Three choices for the
scale factor $\mu$ are considered: the strong dependence of $R_{AA}$ upon this
parameter is quite apparent. We have not tried to fix the value of $\mu$ through 
a best fit to the RHIC data; however, in the following we employ the value
$\mu=3\pi T/2$, the one, among those displayed in Fig.~\ref{fig:RAA_mu}, that
turns out to be in better agreement at large momenta with the PHENIX data on
non-photonic electrons (see below). 
Another important effect on $R_{AA}$ can be appreciated by comparing the
results at RHIC (panel (a)) and LHC (panel (b)) energies. Indeed, as already
mentioned in Sect.~\ref{subsec:ini}, the elementary $Q\bar{Q}$ production cross
sections per binary collision entering into the numerator and denominator of
Eq.~(\ref{eq:RAA}), are not the same, because of nuclear effects.
$R_{AA}$ for charm and bottom is then proportional to the ratio of the
corresponding production cross sections. From Table~\ref{tab:qqbarcs} it clearly
appears that for charm the relative play of shadowing and anti-shadowing is
always giving rise to a quenching factor; for bottom, instead, one gets an
enhancement at RHIC and a quenching at LHC (at $\sqrt{s}=5.5$~TeV).

In Fig.~\ref{fig:RAA_hydro} we display the charm and bottom nuclear modification
factors for a fixed value $\mu=3\pi T/2$ of the scale factor and several
choices for the hydrodynamical scenarios (again for minimum bias collisions at
RHIC and LHC). One can see that the uncertainty associated to the treatment
of the hydrodynamical evolution of the QGP is quite modest compared to the one
due to the heavy-quark interactions (Fig.~\ref{fig:RAA_mu}). Specifically, the
treatment of the expanding fluid as either viscous or ideal has practically no
influence on the heavy-quark propagation.
On the other hand, some sensitivity to the choice of the starting
time of the hydrodynamical evolution is seen to exist. In fact, earlier times 
expose the heavy quarks to higher temperatures and give rise to a more
pronounced quenching of $R_{AA}$. 
At the LHC the ideal scenario with $\tau_0=0.1$~fm appears to be somehow in
contradiction with the trend displayed by the other cases, showing smaller
medium effects for a shorter equilibration time. Note that it might just signal
some shortcomings of the hydrodynamical code in a regime where it has received
little testing. 
Up to semi-peripheral collisions the effect of an earlier equilibration time is
anyway moderate and smaller not only than other theoretical uncertainties, but 
also than the experimental ones. However, as we shall see below, for peripheral
collisions the effect due to the choice of $\tau_0$ turns out to be more 
important.

In Fig.~\ref{fig:RAA_QAe} we display the charm and bottom nuclear modification
factors of heavy quarks, open heavy-flavour hadrons and electrons for viscous
hydrodynamics ($\tau_0=1$~fm), scale factor $\mu=3\pi T/2$ and minimum bias
collisions at RHIC and LHC. The effect of hadronization and decay can be summed
up as a general ``softening'' of the nuclear modification factor, especially
after the decay of B mesons.

In Fig.~\ref{fig:RAA_phenix} the results of our calculations of the nuclear
modification factor of open heavy-flavour electrons at RHIC (for the case of
viscous  
hydrodynamics with $\tau_0=1$~fm and $\mu=3\pi T/2$) are compared to the data
from the PHENIX experiment \cite{phenix,phenix2} in all the centrality
classes. Here we display the $R_{AA}$'s due to the electrons originating either
from $c$-quarks or from $b$-quarks and their combination, showing also the
statistical errors of the calculations. For $p_T\lsim3$~GeV/c the data are
generally underestimated: this is the region where, as already mentioned, the
hadronization mechanism of coalescence has been shown to be important and to
give rise to an enhancement of $R_{AA}$ \cite{rapp}.
At larger momenta the calculations turn out to be on the whole in agreement
with the data from PHENIX, with an evident contribution from the bottom quarks.

\begin{figure*}
\begin{center}
\includegraphics[clip,width=0.8\textwidth]{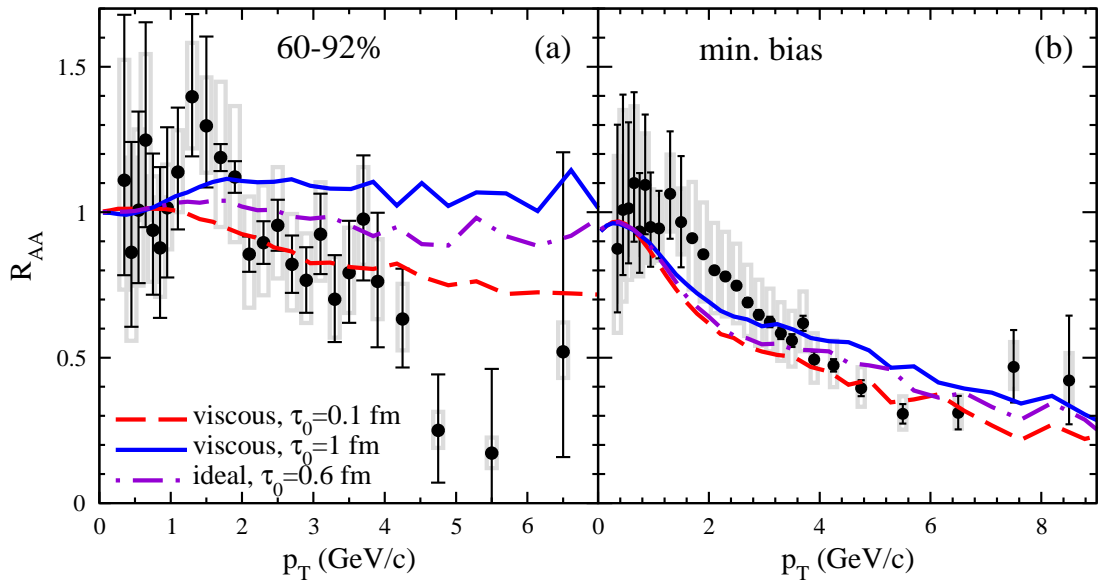}
\caption{(a) The nuclear modification factor of open heavy-flavour electrons
  from 
  peripheral collisions at RHIC for $\mu=3\pi T/2$ and different starting times
  for the hydrodynamical evolution. Data are from the PHENIX experiment
  \protect\cite{phenix,phenix2}. The theoretical curves show (without
  statistical errors) the total yield of electrons from charm and bottom
  quarks. (b) As in panel (a), but for minimum bias collisions.
}
\label{fig:RAA_P60-92}
\end{center}
\end{figure*}

\begin{figure*}
\begin{center}
\includegraphics[clip,width=0.8\textwidth]{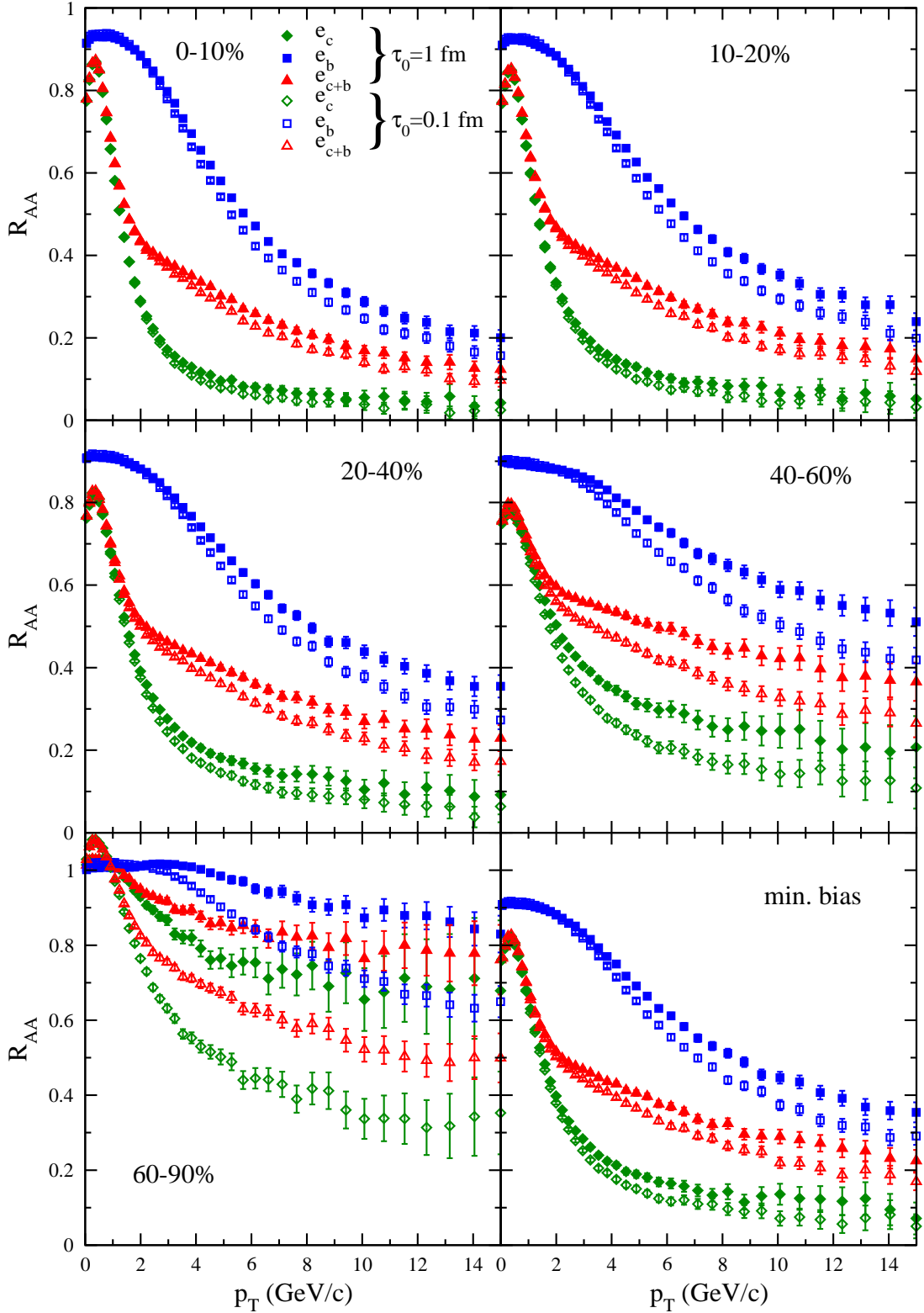}
\caption{The nuclear modification factor of open heavy-flavour electrons at
  LHC@5.5~TeV for $\mu=3\pi T/2$ and viscous hydrodynamics ($\tau_0=1$~fm and
  0.1~fm) in various centrality classes. Electrons originating from charm and
  bottom quarks and their combination are shown, including the statistical
  errors.
}
\label{fig:RAA_LHC5}
\end{center}
\end{figure*}

\begin{figure*}
\begin{center}
\includegraphics[clip,width=0.8\textwidth]{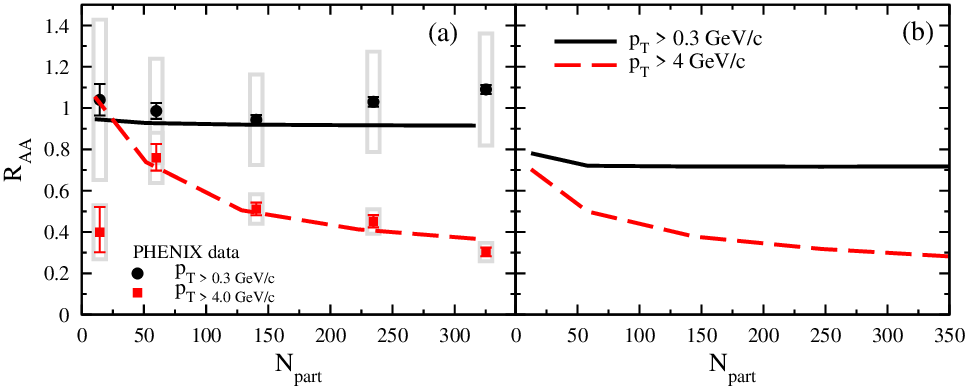}
\caption{The nuclear modification factor of open heavy-flavour electrons at
  RHIC (a) 
  and LHC@5.5~TeV (b) for $\mu=3\pi T/2$ and viscous hydrodynamics
  ($\tau_0=1$~fm), obtained by integrating the electron yields over the
  indicated momentum ranges, as a function of the number of participants. 
  The data points from the PHENIX experiment \cite{phenix2}, including both
  statistical (error bars) and systematic (grey boxes) errors.}
\label{fig:RAA_int}
\vskip 0.2cm
\includegraphics[clip,width=0.8\textwidth]{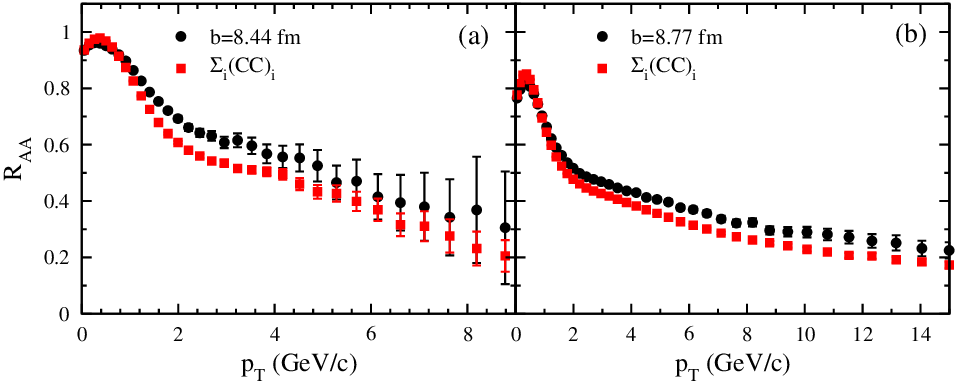}
\caption{The nuclear modification factor of open heavy-flavour electrons at
  RHIC (a) 
  and LHC@5.5~TeV (b) for $\mu=3\pi T/2$ and viscous hydrodynamics
  ($\tau_0=1$~fm). The circles refer to the calculation at a fixed average 
  impact parameter, the squares to the weighted sum of the results in all the
  centrality classes.}
\label{fig:RAA_chkmb}
\end{center}
\end{figure*}

The above considerations are valid in all the centrality classes, except
perhaps the most peripheral one (60-92\%): here the effect of the medium is
expected to be pretty weak and, indeed, our calculation does not show, at low
momenta, any underestimation of the data, but rather it overshoots them at
intermediate momenta. 
In Fig.~\ref{fig:RAA_P60-92}(a) we compare the PHENIX data for the most
peripheral centrality class to the outcome of calculations based on different
initial times. A reduction of $\tau_0$ from 1~fm to 0.1~fm in all the other
centrality classes gives rise to a rather mild effect, well within the
theoretical and experimental uncertainties, as one can see, e.~g., in
Fig.~\ref{fig:RAA_P60-92}(b) for minimum bias collisions. For very peripheral
collisions, on the other hand, the highest temperatures associated to earlier
equilibration times provides a stronger signature. 
Note, however, that for this centrality class --- where one is moving from a
region where the core of the nuclear density distributions is probed to a
region where only the tails of the density distributions are involved --- the
approximation based on the use of an average impact parameter might be less 
reliable.

In Fig.~\ref{fig:RAA_LHC5} we display the nuclear modification factor of open
heavy-flavour electrons at LHC@5.5~TeV for QCD scale factor $\mu=3\pi T/2$ and
two 
choices of the equilibration time. The features of $R_{AA}$ appear to be quite
similar to the case of RHIC, except for a larger quenching at small momenta ---
due in our calculation to the smaller $pp$ over $AA$ ratio of the $c\bar{c}$
production cross sections --- and for a slightly smaller value at large momenta.
The choice of the equilibration time, as already discussed, has a mild effect,
except perhaps on the most peripheral collisions.

In Fig.~\ref{fig:RAA_int} the nuclear modification factor of open 
heavy-flavour electrons, obtained by integrating the electron yields above a
given transverse momentum, is reported as a function of the number of
participants. When the integration range covers most of the spectrum, one
should get the ratio of the total electron yields in $AA$ and $pp$ collisions,
that is, in practice, the ratio of the respective heavy-quark production cross
sections. When only large transverse momenta are selected, one observes the 
quenching of $R_{AA}$ due to the softening of the heavy-quark spectra by the
medium. The calculations are again in fair agreement with the PHENIX data,
except, as already noted, for the most peripheral collisions.
These, however, represent a real puzzle, since they seem to imply that the
strongest medium effects manifest themselves with the highest and the lowest
numbers of participants.

\begin{figure*}
\begin{center}
\includegraphics[clip,width=0.8\textwidth]{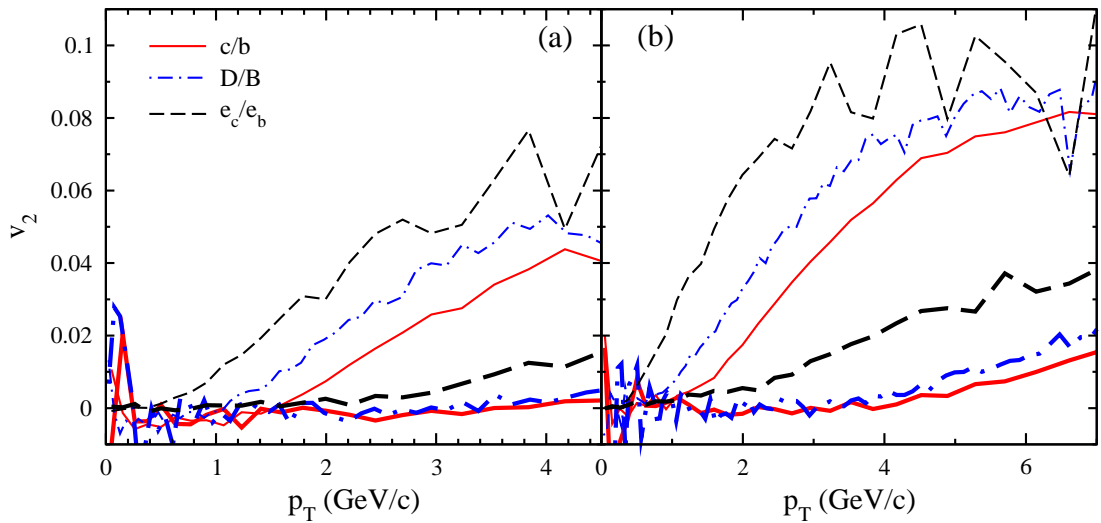}
\caption{(a) The minimum bias anisotropy parameter $v_2$ of heavy quarks,
  open heavy-flavour hadrons and electrons at RHIC for $\mu=3\pi T/2$, viscous
  hydrodynamics ($\tau_0=1$~fm) and $b=8.44$~fm; both the charm (light lines)
  and bottom (heavy lines) sectors are shown.
  (b) As in panel (a), but for LHC at $b=8.77$~fm.}
\label{fig:v2_QAe}
\end{center}
\end{figure*}

\begin{figure*}
\begin{center}
\includegraphics[clip,width=0.8\textwidth]{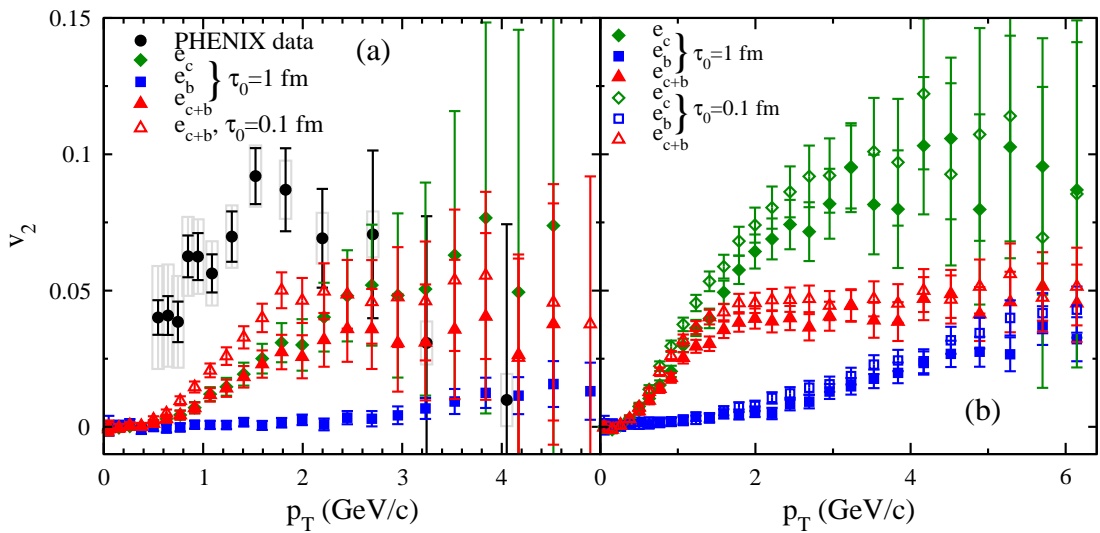}
\caption{(a) The minimum bias anisotropy parameter $v_2$ for open heavy-flavour
  electrons at RHIC for $\mu=3\pi T/2$ and viscous hydrodynamics ($\tau_0=1$~fm
  and 0.1~fm). The circles are data from the PHENIX experiment
  \protect\cite{phenix,phenix2}, including both statistical (error bars) and 
  systematic (grey boxes) errors; the other points (with the statistical errors
  shown) are the outcome of our calculations for electrons originating from
  charm and bottom quarks and their combination.
  (b) As in panel (a), but for LHC and two choices of $\tau_0$.}
\label{fig:v2}
\end{center}
\end{figure*}

Finally, in this study we have also tried to understand the amount of uncertainty
introduced by an approximation that is, to our knowledge, always employed in
this kind of calculations, i.~e. the use of an average impact parameter to
represent a given centrality class. In Fig.~\ref{fig:RAA_chkmb} the minimum bias
nuclear modification factor of open heavy-flavour electrons at RHIC and LHC is
displayed, comparing the results obtained by using a single average impact
parameter to those obtained by combining the weighted spectra in all the
centrality (sub)classes (0-10\%, 10-20\%, ...). The effect is not dramatic
compared to other uncertainties, especially at LHC, but for an accurate
estimate of $R_{AA}$ it will have to be accounted for.

\subsection{Elliptic flow coefficient}

The anisotropy parameter associated to the elliptic flow of the heavy quark is
defined as 
\begin{equation}
  v_2(p_T) = \frac{\int d\phi 
    \frac{\displaystyle d^2N_{AA}}{\displaystyle d\phi dp_T} \cos 2\phi}
    {\int d\phi \frac{\displaystyle d^2N_{AA}}{\displaystyle d\phi dp_T}},
\end{equation}
where $d^2N_{AA}/d\phi dp_T$ is the doubly differential yield of open
heavy-flavour  
electrons, of D and B mesons or of $c$ and $b$ quarks, depending on which
observable one is considering.
 
In Fig.~\ref{fig:v2_QAe} we display the charm and bottom elliptic flow
coefficients of heavy quarks, open heavy-flavour hadrons and electrons for
viscous 
hydrodynamics ($\tau_0=1$~fm), scale factor $\mu=3\pi T/2$ and minimum bias
collisions at RHIC and LHC. In a momentum range up to a few GeV/c the heavy
quark $v_2$ is represented by a growing function of $p_T$. Since, as already
noted, hadronization and decay to electrons produce a softening of the
respective spectra, the net result is a stronger anisotropy of B/D mesons with
respect to $c$/$b$ quarks and an even stronger one for the open heavy-flavour
electrons. At the LHC energy one gets larger elliptic flow coefficients,
especially the bottom contribution.

In Fig.~\ref{fig:v2} we display the total $v_2$ of open heavy-flavour electrons
for minimum bias events at RHIC and LHC. 
The calculation at RHIC energy is compared to the data from PHENIX, which, 
however, are available only up to $p_T\approx4$~GeV/c.
Our results for $\tau_0=1$~fm underestimate the data for $p_T<3$~GeV/c, as they
also do in the case of the nuclear modification factor. The already mentioned
coalescence mechanism for the hadronization process is known to be important
in this momentum region and to provide an enhancement of both $R_{AA}$ and 
$v_2$ \cite{rapp}. Note, however, that an earlier equilibration time gives
rise to an anisotropy parameter closer to the data.
At the LHC energy, one observes a stronger anisotropy for the exclusive $c$ and
$b$ contributions, while the effect due to the choice of $\tau_0$ turns out to
be milder.

\section{Summary and conclusions}
\label{sec:concl}

In this work we intended to provide a consistent study of the stochastic
dynamics of heavy quarks ($c$ and $b$) in the hot environment formed in
high-energy heavy-ion collision experiments. The background medium has been
taken as a fluid (the QGP locally in thermal equilibrium), whose evolution is 
governed by relativistic hydrodynamics, while the interaction of the heavy
quarks with the plasma has been treated within a weakly-coupled framework.

To accomplish this programme one has to evaluate the relevant transport
coefficients 
(accounting, in our case, for the squared momentum acquired per unit time)
characterizing the dynamics of the heavy quarks coupled to the medium.  
We have done this by summing the contributions of soft and hard collisions in 
the plasma. The first ones have been described in the HTL approximation, while 
the latter have been treated within a kinetic pQCD calculation.  
The propagation of the heavy quarks in the expanding plasma has been described
by a relativistic generalization of the Langevin equation.

When the hydrodynamical evolution of the QGP has reached the region of energy
density where the quark to hadron transition may occur, the heavy quarks are
let hadronize and eventually decay into electrons: in fact, their spectra at
RHIC are so far the only source of information on the heavy-flavour dynamics in
the QGP (while waiting for the forthcoming open-charm analyses at RHIC and LHC).
Results have been presented for the invariant single-electron spectra, for the
nuclear modification factor $R_{AA}$ and for the elliptic flow coefficient $v_2$.

The main goal of our work has been first of all to provide of a fully
consistent weakly-coupled calculation (accounting for medium effects at the
lowest non-trivial order), to be viewed as a benchmark for more advanced
studies or less conventional scenarios. 
Nevertheless, the overall degree of agreement with the current experimental data
looks quite satisfactory, suggesting that the actual values of the heavy-quark
transport coefficients might be not too different from what predicted by a weak
coupling approach. 
In particular, for large enough values of $p_T$ (say $p_T\gsim3$~GeV/c), it
has been possible to reproduce the experimental quenching of the single-electron
spectra for any centrality class (except perhaps the most peripheral events,
which will be discussed in more detail in the following). 
On the other hand, at low $p_T$ ($\lsim3$~GeV/c) our results for $R_{AA}$ are
consistently below the data, predicting too much quenching. This outcome might 
be a consequence of the implemented hadronization scheme, where we have
accounted only for the fragmentation mechanism (which entails a degradation of
the heavy-quark momentum). Hadronization through coalescence with light partons
of the medium could help in getting closer to the experimental data. 
Hadronization represents therefore an important source of systematic
uncertainty for  moderate values of $p_T$. 

Concerning the sensitivity to the properties of the background medium, in
general we have found a rather mild dependence on the hydrodynamical scenario
(ideal or viscous, with different choices of $\tau_0$) used to describe the
plasma, except for the most peripheral events, which deserve some separate
comments. 
Quite surprisingly, though with large error bars, experimental data on
heavy-flavour electrons from RHIC display a sizable quenching of the
$p_T$-spectra even in the $60-92\%$ centrality class. 
As shown in Sect.~\ref{subsec:RAA}, employing a value of $\tau_0\sim1$~fm, as
often done in hydrodynamical simulations at RHIC energy, would lead to
$R_{AA}\sim1$, thus essentially showing the absence of medium effects. 
On the other hand, assuming a much faster thermalization, by setting
$\tau_0=0.1$ fm, it is possible to get much closer to the data. 
Notice that with such a small value for the equilibration time it is also
possible to reach a slightly more satisfactory agreement with the elliptic flow
$v_2$ measured for heavy-flavour electrons at RHIC.
At the present stage we cannot draw any definite conclusion on this point and
we leave it as an open issue for future investigation.  

Besides analyzing the RHIC data, we have also provided predictions for LHC
(at the highest energy that should be reached, $\sqrt{s}_{NN}=5.5$~TeV). 
In the absence of experimental data to constrain the properties of the
background medium, we have explored some of the possible hydrodynamical
scenarios proposed in the literature. 
As a general outcome, charm is found to display a stronger quenching and a much
larger elliptic flow with respect to RHIC. Also the spectra of bottom are more
suppressed and characterized by a modest elliptic flow. However, in the combined
electron spectra ($e_{c+b}$), due to the larger relative weight of the $b$
contribution, differences between RHIC and LHC are less evident. 
The possibility by the ongoing experiments of identifying the separate
contribution of charm, by looking at its hadronic decay channels, will
certainly increase the amount of information on the heavy-flavour dynamics
provided by the data.
Heavy-ion collisions at LHC are presently being performed at the NN
center-of-mass energy $\sqrt{s}_{NN}=2.76$~TeV: work is in progress in order to
analyze, within the framework presented in this article, also this kinematical
setup.

\begin{acknowledgement}
We thank R. Averbeck, G. Ridolfi, A. Toia and J. Bielcik for helpful
discussions and P.B. Gossiaux for providing us his data on the transport
coefficients.
\end{acknowledgement}

\appendix

\section{Transverse momentum broadening of heavy quarks in nucleus-nucleus
  collisions} 
\label{app:kTAA}

As explained in the text, our procedure to account for the transverse
momentum broadening in a collinear collision consists in adding, to the heavy
quarks generated in each event, a transverse component randomly chosen
according to a Gaussian distribution~\cite{vogt}
\begin{equation}
  g(k_T) = \frac{1}{\pi\langle k_T^2\rangle}\exp(-k_T^2/\langle k_T^2\rangle).
\label{eq:ktdistr}
\end{equation}
As already mentioned, for $pp$ collision the typical value of 
$\langle k_T^2\rangle_{\rm NN}$ is $1 {\rm ~GeV}^2/{\rm c}^2$.

Let us now consider a collision between the nuclei A and B at impact parameter
$\vec{b}$, in which a $Q\bar{Q}$ pair is produced in a hard event occurring at 
position $\vec{s}$ in the transverse plane (measured with respect to the center
of nucleus A) and at the longitudinal coordinates (measured in the rest frames
of the respective nuclei) $z_A$ and $z_B$. The heavy-quark pair will be 
produced by partons (mainly gluons) that have already crossed a certain
thickness of nuclear matter, thus having acquired some transverse momentum. 
In Refs.~\cite{huf1,huf2} the average squared-momentum of the produced
$Q\bar{Q}$ heavy meson has been found to be given, neglecting absorption
effects, by the expression 
\begin{eqnarray}
  \langle p_T^2\rangle_{AB}^{Q\bar{Q}}(\vec{b},\vec{s},z_A,z_B) &=&
    \langle p_T^2\rangle_{\rm NN}^{Q\bar{Q}} \\
  && + a_{gN}[l_A(\vec{s},z_A) + l_B(\vec{s}-\vec{b},z_B)], \nonumber
\end{eqnarray}
where $\langle p_T^2\rangle_{\rm NN}^{Q\bar{Q}}$ is the value for the same
quantity in NN collisions and  
\begin{eqnarray}
  l_A(\vec{s},z_A) &\equiv& \int_{-\infty}^{z_A}dz\,\rho_A(\vec{s},z)/\rho_0 
    \nonumber \\
  l_B(\vec{s}-\vec{b},z_b) &\equiv& \int_{z_B}^{+\infty}dz\,
    \rho_B(\vec{s}-\vec{b},z)/\rho_0 
\label{eq:lAB}
\end{eqnarray}
represent the lengths traveled by a gluon of B in nucleus A and vice versa.
In Eq.~(\ref{eq:lAB}) $\rho(\vec{s},z)$ is the nuclear density function and 
$\rho_0$ is the central nuclear density.
The parameter $a_{gN}$ represents the average squared (transverse) momentum per
unit length acquired by a gluon in nuclear matter. At the energy of SPS one
finds good agreement with the $J/\psi$ data for $a_{gN}\approx 0.08$~GeV$^2$/fm
\cite{abreu}. 

Let us now apply the above scheme, developed for charmonia, to the problem of
inclusive single-particle spectra. The random transverse momentum kick of the
pair reflects the one acquired by the incoming gluons in crossing nuclear
matter. It is shared by the two quarks,
\begin{equation}
  \vec{k}_T^{Q\bar{Q}} = \vec{k}_T^Q+\vec{k}_T^{\bar{Q}},
\end{equation}
and for a random choice of the azimuthal angle of each quark of the pair one
has: 
\begin{equation}
  \langle \vec{k}_T^2\rangle^{Q\bar{Q}} = \langle \vec{k}_T^2\rangle^Q +
    \langle \vec{k}_T^2\rangle^{\bar{Q}} + 2 \langle k_T^Q k_T^{\bar{Q}}
    \cos\phi_{Q\bar{Q}}\rangle \equiv 2 \langle \vec{k}_T^2\rangle^Q. 
\end{equation}
In general one does not know the longitudinal coordinate of the hard event, so
that one takes an average over the longitudinal positions $z_A$ and $z_B$,
with a weight given by the corresponding local nuclear density. The average 
squared momentum used in the distribution of Eq.~(\ref{eq:ktdistr}) to extract
the transverse momentum kick for the heavy quarks is then: 
\begin{eqnarray}
  &&\langle k_T^2\rangle_{AB}^Q(\vec{b},\vec{s}) = \langle 
    k_T^2\rangle_{\rm NN}^Q \nonumber \\
  && \qquad + \frac{a_{gN}}{2}\left[\frac{
    \int_{-\infty}^{+\infty}dz_A\,\rho_A(\vec{s},z_A) l_A(\vec{s},z_A)} 
    {\int_{-\infty}^{+\infty}dz_A\,\rho_A(\vec{s},z_A)}+\right.
    \nonumber \\
  && \qquad
    \left.+\frac{\int_{-\infty}^{+\infty}dz_B\,\rho_B(\vec{s}-\vec{b},z_B)
    l_B(\vec{s}-\vec{b},z_B)} 
    {\int_{-\infty}^{+\infty}dz_B\,\rho_B(\vec{s}-\vec{b},z_B)} \right].
\label{eq:kT2AB}
\end{eqnarray}
Note that we keep track of the transverse position of the hard event, since the 
transverse momentum kick is imparted to heavy quarks distributed in the
transverse plane according to binary-collision scaling. 

The parameter $a_{gN}$ has been fixed in Refs.~\cite{huf1,huf2} at an energy
lower that the ones we are interested in ($\sqrt{s}=158$~GeV \cite{abreu}).
In order to extrapolate its value to the RHIC and LHC regimes we have
considered the transverse squared momentum of Eq.~(\ref{eq:kT2AB}) averaged
over the $Q\bar{Q}$ pair transverse position $\vec{s}$ and over the impact
parameter $\vec{b}$. This same quantity has been modeled in
Ref.~\cite{mang,vogt} in terms of the squared-momentum transfer per collision,
\begin{equation}
  \Delta^2(\mu)=0.225\frac{\ln^2(\mu/{\rm GeV})}{1+\ln(\mu/{\rm GeV})}\,{\rm
  GeV}^2, 
\end{equation}
($\mu\cong2m_Q$ being the interaction scale) and of the inelastic
nucleon-nucleon cross section $\sigma_{\rm NN}$. A straightforward comparison
of our expression with the one of Ref.~\cite{vogt} yields, indeed, 
\begin{equation}
  a_{gN} = \Delta^2\sigma_{\rm NN}\rho_0,
\end{equation}
$\sigma_{\rm NN}$ being the total inelastic NN cross section and $\rho_0$ the
central nuclear density, and leads to the values displayed in
Table~\ref{tab:agN}.
\begin{table}
\caption{\label{tab:agN} The total nucleon-nucleon inelastic cross section and
  the parameter $a_{gN}^{Q\bar{Q}}$ at different kinematics.
}
\begin{center}
\begin{tabular}{|c|ccc|}
\hline
 $\sqrt{s}$ (GeV)                & 158   & 200  & 5500 \\ 
\hline
  $\sigma_{\rm NN}$ (mb)         & 33    & 42   & 72   \\
\hline
  $a_{gN}^{c\bar{c}}$ (GeV$^2/$fm) & 0.081 & 0.10 & 0.17 \\
\hline
  $a_{gN}^{b\bar{b}}$ (GeV$^2/$fm) & 0.221 & 0.27 & 0.47 \\
\hline
\end{tabular}
\end{center}
\end{table}

\section{Momentum diffusion coefficients: calculation of the hard
  contribution}
\label{app:hard}
 
In this Appendix we provide some details on the calculation of the
hard-collision contribution to the momentum-diffusion coefficients. 
The squared amplitude for scattering on quarks (Fig.~\ref{fig:quarkgluon}) 
reads:
\begin{eqnarray}
  \left|\overline{{\cal M}}_q(s,t)\right|^2 &=& 2N_f 2N_c \alpha_s^2\pi^2 
    \frac{64}{9} \\
   && \times \frac{(M^2-u)^2+(s-M^2)^2+2M^2t}{t^2}, \nonumber
\end{eqnarray}
where, as mentioned in the text, the factor $2N_f$ arises from the equal
contribution of all light quark and anti-quark flavours, while the further
factor $2N_c$ with respect to Ref.~\cite{com} reflects our summing, rather than
averaging, over the helicities and colours of the incoming light quark. 
For scattering on gluons (Fig.~\ref{fig:quarkgluon}) one gets
\begin{eqnarray}
  &&\left|\overline{{\cal M}}_g(s,t)\right|^2 = 2(N_c^2-1) \alpha_s^2\pi^2
    \nonumber \\
  && \quad \times \left[32 \frac{(s-M^2)(M^2-u)}{t^2}\right. \nonumber \\
  && \qquad + \frac{64}{9} \frac{(s-M^2)(M^2-u)+2M^2(s+M^2)}{(s-M^2)^2}
      \nonumber \\
  && \qquad + \frac{64}{9} \frac{(s-M^2)(M^2-u)+2M^2(M^2+u)}{(M^2-u)^2}
      \nonumber \\
  && \qquad + \frac{16}{9} \frac{M^2(4M^2-t)}{(s-M^2)(M^2-u)}
      \nonumber \\
  && \qquad  + 16\frac{(s-M^2)(M^2-u)+M^2(s-u)}{t(s-M^2)} \nonumber \\
  && \qquad \left. -16\frac{(s-M^2)(M^2-u)-M^2(s-u)}{t(M^2-u)}\right],
\end{eqnarray}
where the factor $2(N_c^2-1)$ with respect to Ref. \cite{com} stems from
summing over the polarization and the colours of the incoming gluon. 

In order to perform the phase space integration in Eqs.~(\ref{eq:kthard}) and
(\ref{eq:klhard}), it is convenient to express the longitudinal and transverse
squared momenta exchanged in the collision in terms of the energy $\omega$
lost by the heavy quark and of the Mandelstam variable $t$. One has
\begin{equation}
  q_L^2 = \frac{(2E\omega+|t|)^2}{4p^2} \quad {\rm and} \quad
  q_T^2 = \omega^2+|t|-\frac{(2E\omega+|t|)^2}{4p^2},
\end{equation}
so that one can follow the procedure of Ref.~\cite{pei1} for the evaluation of
integrals like ($\int_k\equiv \int d\vec{k}/(2\pi)^3$)
\begin{eqnarray}
  I[f] &=& \frac{1}{2E}\int_k\frac{n_{B/F}(k)}{2k} \int_{k'}
    \frac{1\pm n_{B/F}(k')}{2k'}\int_{p'}\frac{\theta(|t|-|t|^*)}{2E'}
    \nonumber \\ 
  && \times(2\pi)^4\delta^{(4)}(P+K-P'-K')f(s,t,\omega).
\end{eqnarray}
One gets
\begin{eqnarray}
  I[f] &=& \frac{1}{16\pi^2 p E} \int_k\frac{n_{B/F}(k)}{2k}
    \int_{|t|^*}^{|t|^{\rm max}}d|t| \int_{\omega_{\rm min}}^{\omega_{\rm max}}
    \frac{d\omega}{\sqrt{g(\omega)}} \nonumber \\
  && \times [1\pm n_{B/F}(k+\omega)]f(s,t,\omega), 
\end{eqnarray}
where
\begin{equation}
  g(\omega) = -a^2\omega^2+b\omega+c,
\end{equation}
being
\begin{eqnarray}
  a &=& \frac{s-M^2}{p},\quad b=\frac{2|t|}{p^2}\left[E(s-M^2)-k(s+M^2)\right],
    \\
  c &=& \frac{|t|}{p^2}\left[4p^2k^2-(s-M^2-2Ek)-|t|((E+k)^2-s)\right] 
    \; \nonumber
\end{eqnarray}
and
\begin{equation}
  \omega_{\rm max/min}=\frac{b\pm\sqrt{b^2+4a^2c}}{2a^2}.
\end{equation}
From $b^2+4a^2c\ge 0$ one gets then:
\begin{equation}
|t|^{\rm max}=\frac{(s-M^2)^2}{s}.
\end{equation}
Hard collisions provide a non-vanishing contribution only as long as $|t|^{\rm
  max}>|t|^*$. This allows to set the relevant range of integration over
$\vec{k}$, which is given by: 
\begin{eqnarray}
  && k>\frac{|t|^*}{4E(1+v)}\left(1+\sqrt{1+\frac{4M^2}{|t|^*}}\right), \\
  && -1<\cos\theta_{kp}<{\rm min}\left[1,\frac{1}{v}-\frac{|t|^*}{4Ekv}
    \left(1+\sqrt{1+\frac{4M^2}{|t|^*}}\right)\right]. \nonumber
\end{eqnarray}

\end{document}